\documentclass[12pt,letterpaper]{article}
\pdfoutput=1

\usepackage{color}
\definecolor{refkey}{rgb}{1,0,0}
\definecolor{labelkey}{rgb}{0,0,1}
\usepackage{amsfonts,amssymb,amsmath}
\usepackage{MnSymbol,wasysym}
\usepackage{esint}

\usepackage{graphicx}
\usepackage{subfig}
\usepackage{caption}
\usepackage{booktabs}

\usepackage{listings}

\usepackage{hyperref}
\hypersetup{
	colorlinks=true,
	citecolor=blue,
    linkcolor=blue,
    filecolor=magenta,      
    urlcolor=cyan,
}
\numberwithin{equation}{section}

\newcommand{\be}{\begin{equation}}
\newcommand{\ee}{\end{equation}}
\newcommand{\ben}{\begin{displaymath}}
\newcommand{\een}{\end{displaymath}}
\newcommand{\bea}{\begin{eqnarray}}
\newcommand{\eea}{\end{eqnarray}}
\newcommand{\bean}{\begin{eqnarray*}}
\newcommand{\eean}{\end{eqnarray*}}

\def\a {\alpha}

\newcommand{\bra}[1]{\mbox{$\langle #1 |$}}
\newcommand{\ket}[1]{\mbox{$| #1 \rangle$}}

\newcommand{\eg}{{\it e.g.}}
\newcommand{\ie}{{\it i.e.}}

\newcommand{\tr}{\mbox{Tr}}

\newcommand{\commentout}[1]{}

\usepackage{amsmath}

\newcommand{\beq}{\begin{equation}}
\newcommand{\eeq}{\end{equation}}
\newcommand{\beqr}{\begin{displaymath}}
\newcommand{\eeqr}{\end{displaymath}}
\newcommand{\beqa}{\begin{eqnarray}}
\newcommand{\eeqa}{\end{eqnarray}}
\newcommand{\beqar}{\begin{eqnarray*}}
\newcommand{\eeqar}{\end{eqnarray*}}

\newcommand{\non}{\nonumber}

\newcommand{\cF}{{\cal F}}

\newcommand{\cC}{{\cal C}}

\newcommand{\half}{\ensuremath{\frac{1}{2}}}

\renewcommand{\Re}{\ensuremath{\mathrm{Re}}}
\renewcommand{\Im}{\ensuremath{\mathrm{Im}}}

\begin{document}

\title{\LARGE \bf The R-matrix bootstrap for the 2d O(N) bosonic model with a boundary}

\author{
	Martin Kruczenski$^{1,2}$\,\thanks{E-mail: \texttt{markru@purdue.edu, muralih@purdue.edu}} \ and
	Harish Murali$^1$\\
	$^1$ Dep. of Physics and Astronomy, and \\
	$^2$ Purdue Quantum Science and Engineering Institute (PQSEI), \\ 
	Purdue University, W. Lafayette, IN  }

\maketitle

\begin{abstract}
 The S-matrix bootstrap is extended to a 1+1d theory with $O(N)$ symmetry and a boundary in what we call the R-matrix bootstrap since the quantity of interest is the reflection matrix (R-matrix). Given a bulk S-matrix, the space of allowed R-matrices is an infinite dimensional convex space from which we plot a two dimensional section given by a convex domain on a 2d plane.  In certain cases, at the boundary of the domain, we find vertices corresponding to integrable R-matrices with no free parameters. In other cases, when there is a one-parameter family of integrable R-matrices, the whole boundary represents integrable theories. We also consider R-matrices which are analytic in an extended region beyond the physical cuts, thus forbidding poles (resonances) in that region. In certain models, this drastically reduces the allowed space of R-matrices leading to new vertices that again correspond to integrable theories. We also work out the dual problem, in particular in the case of extended analyticity, the dual function has cuts on the physical line whenever unitarity is saturated. For the periodic Yang-Baxter solution that has zero transmission, we computed the R-matrix initially using the bootstrap and then derived its previously unknown analytic form.
\end{abstract}

\clearpage
\newpage



\section{Introduction}
\label{intro}
\subsection{The S-matrix and R-matrix bootstrap programs}

  New insights were recently found on the old idea \cite{Smatrix} of determining the S-matrix directly from its analytic structure, symmetries, crossing, and unitarity. This certainly works in two dimensional integrable theories but only after using the factorization constraint, namely the Yang-Baxter equation. Without that, those constraints are not enough to completely determine the S-matrix. However, recently it was found that maximizing the coupling between particles and their bound states led to well-known theories such as a subsector of the sine-Gordon model. It can be also applied to 3+1 dimensional theories, and multiple amplitudes \cite{Paulos:2016fap,Paulos:2016but,Paulos:2017fhb,Homrich:2019cbt}. The main physical argument is that, when the spectrum of bound states is fixed, there is a limit on the value of the coupling since stronger couplings will lead to more bound states. This is a very powerful idea, namely that certain theories lay at particular points of the space of allowed theories (or S-matrices) and that those particular points can be found by maximizing certain functionals in that space. For this paper, the case of interest is the 2d $O(N)$ non-linear sigma model studied in \cite{ONmodel}, exactly solved in \cite{Zamolodchikov:1978xm} and more recently revisited with the S-matrix bootstrap approach in  \cite{He:2018uxa,Cordova:2018uop,Paulos:2018fym}. In particular, in \cite{He:2018uxa} it was argued that maximizing a linear functional in a convex space generically leads to vertices of the space. In fact, it was shown that the $O(N)$ non-linear sigma model (NLSM) lies at one of those vertices, the functional just being a way to find it. Later this was made more manifest in \cite{Cordova:2019lot} where a section of the space was plotted with a clear vertex at the NLSM. However, it was also found that other theories did not appear to be at vertices. Further work on other models \cite{Bercini:2019vme} showed that sometimes full regions of the boundary correspond to interesting theories if such theories have free parameters. Various other ideas have been discussed in the context of the S-matrix bootstrap and similar methods applied to gapped theories\cite{Guerrieri:2020bto, Guerrieri:2020kcs, EliasMiro:2019kyf, Guerrieri:2018uew, Hebbar:2020ukp,Behan:2020nsf,Huang:2020nqy,Komatsu:2020sag,Elvang:2020lue,Bose:2020shm,Correia:2020xtr,Karateev:2019ymz,Nayak:2017qru,Sever:2017ylk,Doroud:2018szp,Behan:2018hfx,Anderson:2016rcw,Chim:1995kf}.
 
  Motivated by this, we consider the two dimensional bosonic $O(N)$ model with a boundary \cite{Ghoshal:1994bc,Moriconi:1998gc,Moriconi:2001xz,Aniceto:2017jor} and compute the reflection matrix (R-matrix). The procedure is similar. We impose all constraints of analyticity, crossing, and unitarity and map the allowed space of R-matrices looking for special points at the boundary of the space. A difference is that crossing depends on the bulk S-matrix that we have to specify initially. It might be interesting to find the S-matrix and R-matrix simultaneously, but the constraints are non-linear. In fact, it seems more straightforward to compute first the S-matrix and then the R-matrix. In any case, this is the procedure we follow here and find the same variety of phenomena previously discussed for the S-matrix. When an integrable R-matrix exists with no free parameters it usually appears at a vertex of the allowed space. If this is not the case,  we apply a new procedure that we call extended analyticity where we extend the analytic properties of the functions beyond the physical cuts. This severely restricts the allowed space by eliminating the R-matrices that have poles in that extended region. This is similar to removing R-matrices with bound states, but in this case, we can say that we remove resonances\footnote{Generically speaking, since some poles are too far from the real axis to be considered well-defined resonances.}. We find that R-matrices that were not at vertices now appear at the vertices of the restricted space. In other cases, there is a one parameter family of integrable R-matrices. In that case, we find that all the boundary corresponds to integrable R-matrices. We also find vertices that do not seem to correspond to any known theory.
  
  The paper is organized as follows: In the rest of this section, we describe the properties of the S-matrix and R-matrix and exact results that follow from integrability with several examples. One result is new, we obtain an integrable reflection matrix for the periodic Yang-Baxter solution with no transmission (pYB). In the following section, we forget the requirement of integrability and just map out the allowed space of R-matrices from generic constraints. There we find the aforementioned properties. In the next section, we discuss the dual problem and find some useful properties of the problem with extended analyticity. In particular, we argue that it has to be regularized and that now, unitarity saturation does not follow automatically. In spite of that, all the R-matrices we found numerically actually saturate unitarity. In the last section, we give our conclusions.  
 
\subsection{The 2d O(N) bosonic model, general properties and exact S-matrices} 

 Consider a two dimensional theory with $O(N)$ symmetry with N-species of bosonic particles with equal mass $m$ labeled by $a=1\ldots N$ and two particle scattering $(p_1,a)+(p_2,b)\rightarrow (p_3,c)+(p_4,d)$ given by a generic S-matrix of the form
\begin{equation}
 S_{ab\rightarrow cd}= S_{ab}^{cd}(s)\   \delta(p_1-p_3) \delta(p_2-p_4) +(p_3\leftrightarrow p_4)(c\leftrightarrow d)\ ,
  \label{a1a}
 \end{equation}
 where $s=(p_1-p_2)^2$ is a Mandelstam variable and 
\beq
 S_{ab}^{cd}(s)=\delta_{ab}\delta_{cd}\ S_A(s)+\delta_{ac}\delta_{bd}\ S_T(s) + \delta_{ad}\delta_{bc}\ S_R(s)\ ,
 \label{a1}
\eeq
 where $S_T(s)$, $S_R(s)$ and $S_A(s)$ represent the transmission, reflection and annihilation amplitudes. Equivalently we can write
\begin{equation}
\begin{aligned}
 S_{ab}^{cd}(s)=&
 \frac{1}{N}\delta_{ab}\delta_{cd}S_I(s)+\half(\delta_{ac}\delta_{bd}+ \delta_{bc}\delta_{ad}-\frac{2}{N}\delta_{ab}\delta_{cd}) S_+(s)  \\
 & +\half(\delta_{ac}\delta_{bd}- \delta_{ad}\delta_{bc})S_-(s)\ ,
\end{aligned}
 \label{a2}
\end{equation}
with
\beq
S_I = N S_A + S_T + S_R, \ \ \ S_\pm = S_T\pm S_R. 
 \label{a3}
\eeq
The functions $S_I(s)$ and $S_\pm(s)$ represent the scattering amplitudes in the three isospin channels: isoscalar, symmetric and antisymmetric.
It is convenient to use the relative rapidity variable $\theta$ defined through
\beq
 s = 4 m^2 \cosh^2\frac{\theta}{2}\ ,  
 \label{stheta}
\eeq
If there are no bound states, the functions $S_a(\theta)$ are analytic in the physical strip defined by $0\le \Im \theta \le \pi$. On the real line, $\theta\in\mathbb{R}$ unitarity implies $|S_I(\theta)|\le 1$,\ $|S_-(\theta)|\le 1$, \text{and }$|S_+(\theta)|\le 1$. Finally crossing implies 
\beq
 S_A(i\pi-\theta) = \sum_B C_{AB} S_B(\theta), \ \ \ \ \ A,B = I,+,- ,
 \label{a4}
\eeq
 In this formula, the indices $A,B,\ldots$ label isospin channels and the matrix $C$ satisfies $C^2=1$ and is given by
\beq\label{ipmcrossing}
C = 
 \left(\begin{array}{ccc} \frac{1}{N} & \frac{N}{2}+\half-\frac{1}{N} &\half-\frac{N}{2}\\ 
 	                                  \frac{1}{N} & \half -\frac{1}{N}                   & \half  \\ 
 	                                 -\frac{1}{N} & \half+\frac{1}{N}                   & \half  \end{array}\right)  .
\eeq
Up to now we have only described general constraints on the S-matrix due to standard properties of the field theory. If the theory is integrable the S-matrix should also satisfy the Yang-Baxter equation
\beq
S_{a_1a_2}^{c_1c_2}(\theta)\, S_{c_1a_3}^{b_1c_3}(\theta+\theta')\, S_{c_2c_3}^{b_2b_3}(\theta') = S_{a_2a_3}^{c_2c_3}(\theta')\, S_{a_1c_3}^{c_1b_3}(\theta+\theta')\, S_{c_1c_2}^{b_1b_2}(\theta) \ ,
\label{YBeq}
\eeq
The simplest integrable model is just a free theory whose reflection matrix we briefly discuss in appendix \ref{free_theory_appendix}. In the  main body of this paper we consider two non-trivial  integrable models. One is the 2d $O(N)$ non-linear sigma model solved by Zamolodchikov and Zamolodchikov \cite{Zamolodchikov:1978xm} with S-matrices given by
\beqa
S^{NLSM}_+(\theta)  &=& \frac{\theta-i\lambda}{\theta+i\lambda}\frac{\theta-i\pi}{\theta+i\pi} S^{NLSM}_I(\theta)  \ ,\\
S^{NLSM}_-(\theta) &=& \frac{\theta-i\pi}{\theta+i\pi} S^{NLSM}_I(\theta) \ , \\
 S^{NLSM}_I(\theta) &=&-F_{\pi+\lambda}(\theta)F_{2\pi}(\theta) \ ,
  \label{a5}
\eeqa
where
\beq
\lambda = \frac{2\pi}{N-2}\ ,
 \label{a6}
\eeq
and we introduced the function $F_a(\theta)$ (see \cite{Cordova:2019lot}): 
\beq
F_a(\theta) = \frac{\Gamma\left(\frac{a+i\theta}{2\pi}\right)\Gamma\left(\frac{a-i\theta}{2\pi}+\half\right)}{\Gamma\left(\frac{a-i\theta}{2\pi}\right)\Gamma\left(\frac{a+i\theta}{2\pi}+\half\right)}\ .
 \label{a7}
\eeq
The other is the periodic Yang-Baxter solution (pYB) given by \cite{Hortacsu:1979pu,Cordova:2018uop}
\beqa
 S^{pYB}_+(\theta) &=& H_{\nu/2}\left(\nu+\frac{i\theta\nu}{\pi},2\nu-\frac{i\theta\nu}{\pi};\nu-\frac{i\theta\nu}{\pi},2\nu+\frac{i\theta\nu}{\pi}\right) \ ,\\
 S^{pYB}_-(\theta) &=& -S^{pYB}_+(\theta) \ ,\\
 S^{pYB}_I(\theta) &=& \frac{\sinh\left(\nu(1-\frac{i\theta}{\pi})\right)}{\sinh\left(\nu(1+\frac{i\theta}{\pi})\right)} S^{YB}_+(\theta)\ ,
\label{pYB}
\eeqa
with
\beq
 H_\nu(\alpha,\beta;\gamma,\delta) = \lim_{M\rightarrow\infty} \prod_{|n|\le M} \frac{\Gamma(\frac{\gamma+i\pi n}{4\nu})\Gamma(\frac{\delta+i\pi n}{4\nu})}{\Gamma(\frac{\alpha+i\pi n}{4\nu})\Gamma(\frac{\beta+i\pi n}{4\nu})}\ . \\
 \label{a8}
\eeq
The function $H_\nu$ is studied in more detail in appendix \ref{H_nu}, including an easier to evaluate definition and various properties needed to check unitarity, crossing, and the YB equation. It will appear again in the computation of the R-matrix.
 In \cite{He:2018uxa,Cordova:2018uop,Paulos:2018fym} the S-matrix bootstrap procedure was applied to the $O(N)$ symmetric case with no bound states. In \cite{He:2018uxa} it was observed that the NLSM appears at a vertex of the boundary space, a fact made manifest in \cite{Cordova:2019lot}. In \cite{Cordova:2018uop} the pYB solution was found and seen to correspond to some earlier work \cite{Hortacsu:1979pu}.

\subsection{The 2d O(N) bosonic model, reflection matrices}
\begin{figure}
	\centering
	\includegraphics[width=.9\textwidth]{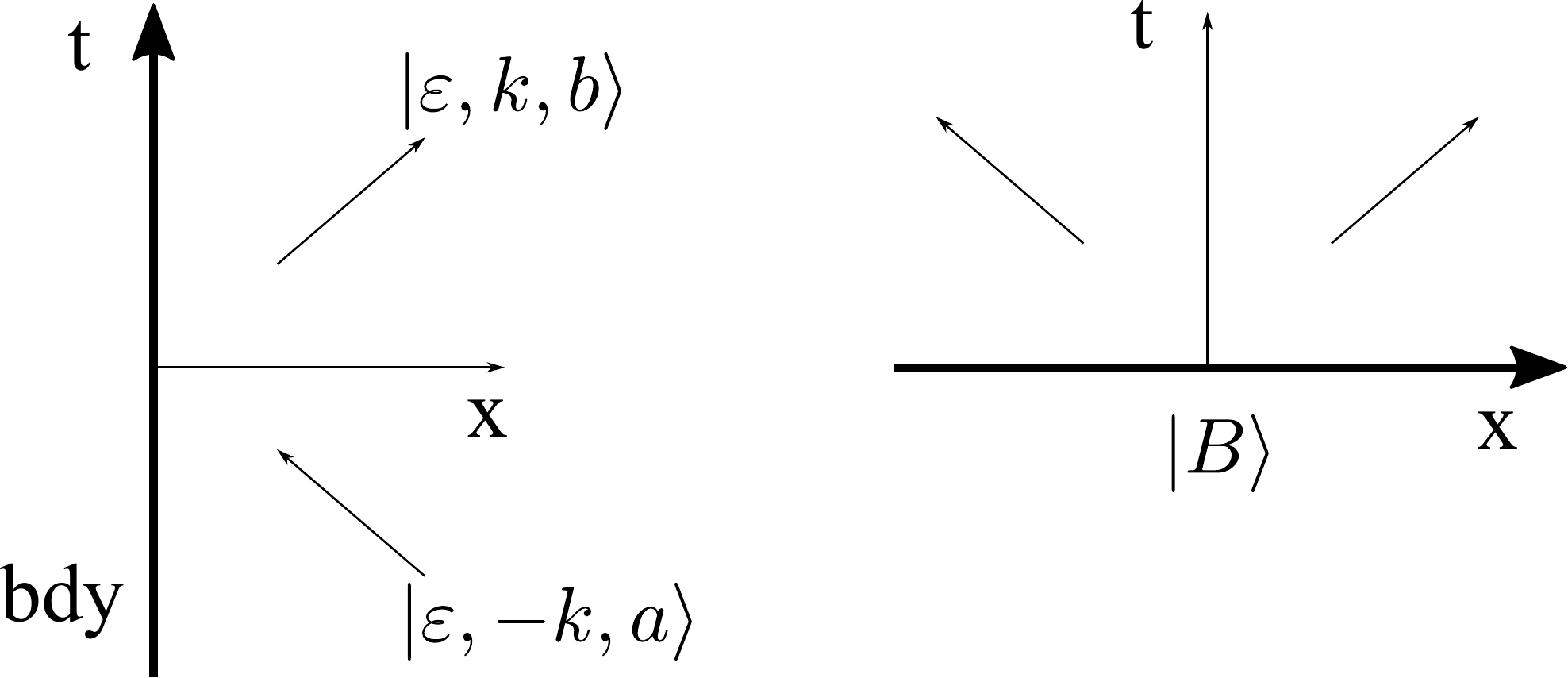}
	\caption{Pictorial description of the $1\rightarrow 1$ R-matrix describing the amplitude for a particle to bounce from the wall possibly changing its identity $a\rightarrow b$. A double Wick rotation relates this process to pair production from an initial boundary state $\ket{B}$. In the second case we can define a bulk S-matrix using the usual asymptotic states.}
	\label{reflection}
\end{figure}

Quantum field theories on a half-line show up in many areas of theoretical physics like D-branes in string theory and defects in condensed matter systems. The main object of interest is the reflection matrix giving the reflection amplitude for a given initial incoming state to be reflected into an outgoing one (see fig.\ref{reflection}). Since the boundary breaks Lorentz invariance, the reflection matrix is a natural function of the incoming energy which, for a single incoming particle can be parameterized in terms of the rapidity $\theta$ as
\beq
 \varepsilon = m \cosh\theta\ .
 \label{a9}
\eeq
The incoming momentum is $k=-m\sinh\theta$. If the outgoing state is a single particle state it will have the same energy but opposite momentum. Taking into account that the theory has $N$ species of particles of mass $m$ labeled by indices $a=1\ldots N$ the $1\rightarrow 1$ R-matrix can be formally defined as
\beq
 R_a^b(\theta) =\ _{\mathrm{out}}\bra{\varepsilon,k,b} \varepsilon,-k,a\rangle_{\mathrm{in}}\ .
 \label{a10}
\eeq
 Allowing for particle production, unitarity requires that
\beq
 |R_a^b(\theta)|\le 1, \ \ \theta\in\mathbb{R}\ .
 \label{a11}
\eeq
\begin{figure}
	\centering
	\includegraphics[width=.8\textwidth]{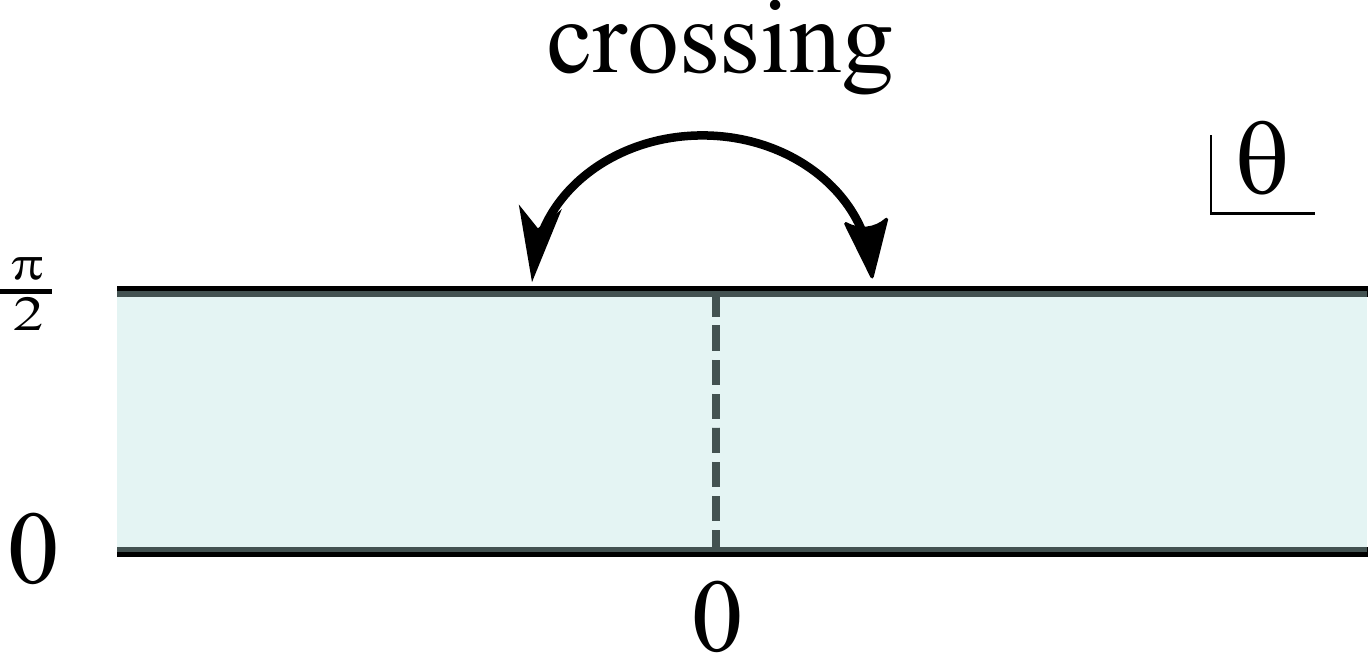}
	\caption{Physical region of the reflection process ($\Re\, \varepsilon \ge 0$). The positive real axis corresponds to the physical reflection ($\varepsilon \in \mathbb{R}_{\ge m}$) which is the boundary value of an analytic function in the physical region (shaded). Crossing can be imposed on the upper line, this is the minimal region where crossing gives a constraint. Boundary bound states could appear as poles on the imaginary axis (dashed line) but we do not allow them here. }
	\label{theta_region}
\end{figure}
\begin{figure}
	\centering
	\includegraphics[width=.75\textwidth]{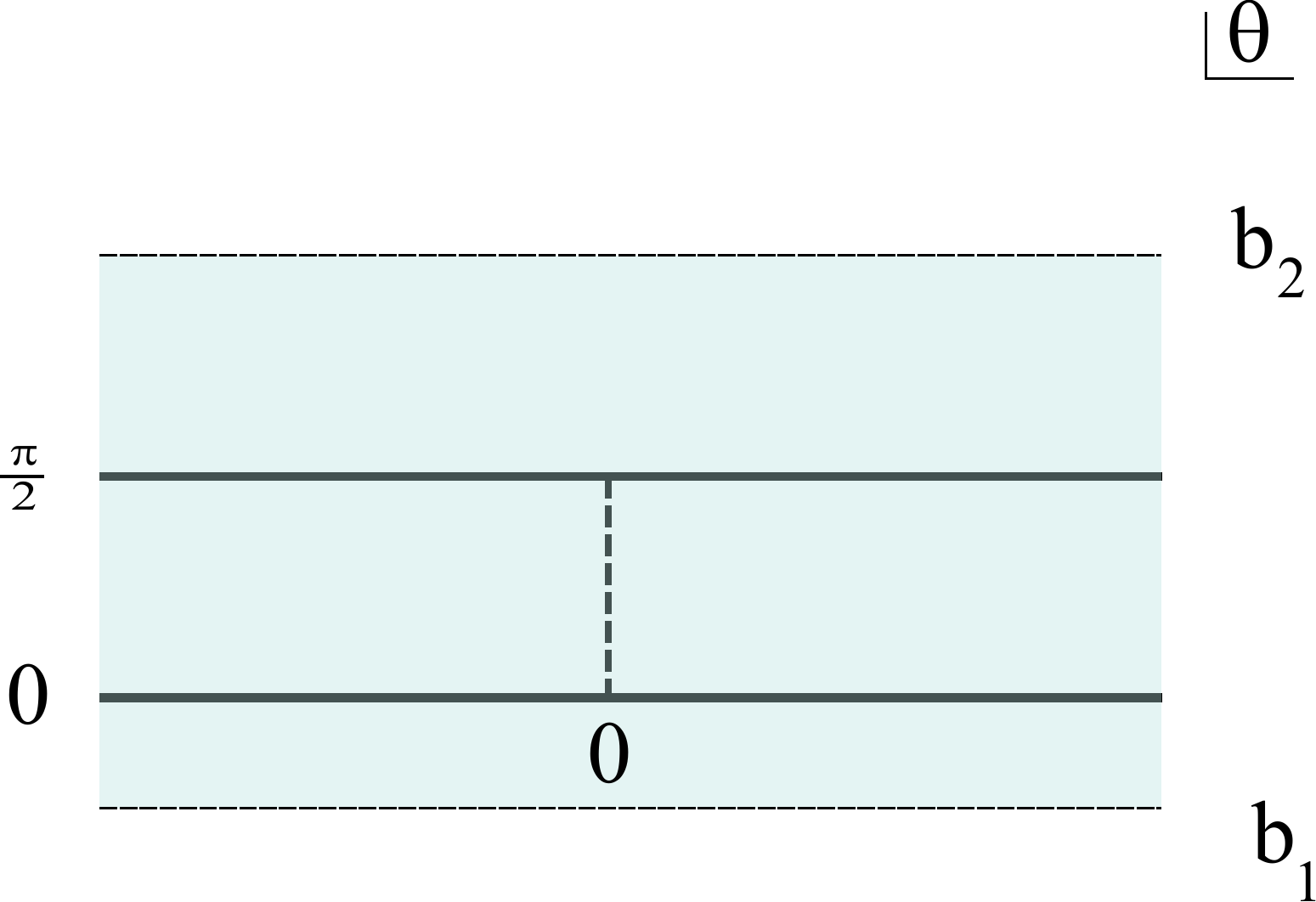}
	\caption{Extended region of analyticity. We considered two cases: $b_1=0$, $\frac{\pi}{2}\le b_2\le \pi$ and $b_1=\pi-b_2$, $b_2>\pi$. It might seem that in the first case analyticity is guaranteed by crossing but this is not so because of the non-trivial crossing equation (\ref{cross_unit}).}
	\label{theta_extended_region}
\end{figure}
The variable $\theta$ can be analytically continued to the strip $0\le\Im \theta\le \frac{\pi}{2}$ (see fig.\ref{theta_region}) or equivalently $\Re\, \varepsilon\ge0$. Poles on the real axis of $\varepsilon$ in the segment $0\le \Re\, \varepsilon \le m$ correspond to bound states of the particle to the boundary. We are going to assume no such bound states exist. For $\varepsilon \ge m$ there is a cut corresponding to intermediate single particle states. Further along the real axis we can have multi-particle cuts as in the case of the S-matrix. A double Wick rotation relates the imaginary axis of the energy plane to the real axis of a situation where the boundary is at $t=0$ and therefore becomes an initial state. The spatial axis would now run from $-\infty$ to $+\infty$ allowing the definition of the usual S-matrix. The $1\rightarrow 1$ reflection process is replaced by particle pair production from the initial state. Using this idea,  Ghoshal and Zamolodchikov \cite{Ghoshal:1993tm} showed that the R-matrix satisfies the following crossing equation for an integrable bulk S-matrix, 
\beq
R_a^b(\frac{i\pi}{2}-\theta) = S_{cd}^{ab}(2\theta)\, R_d^c(\frac{i\pi}{2}+\theta)\ ,
\label{cross_unit}
\eeq
where $S_{cd}^{ab}(\theta)$ is the (flavor part) bulk S-matrix in eq.(\ref{a1}) for the model under consideration. We are still allowing for particle production in the reflection process. Even when no particles are produced in a reflection, the R-matrix generically breaks integrability. The final condition for integrability is the reflection Yang-Baxter equation \cite{Ghoshal:1993tm}: 
\beq
 S^{c_1c_2}_{a_1a_2}(\theta_1-\theta_2) R^{c_3}_{c_1}(\theta_1) S^{c_4a_3}_{c_2c_3}(\theta_1+\theta_2) R^{a_4}_{c_4}(\theta_2) = R^{c_3}_{a_2}(\theta_2)S^{c_1c_2}_{a_1c_3}(\theta_1+\theta_2)R^{c_4}_{c_1}(\theta_1)S^{a_4a_3}_{c_2c_4}(\theta_1-\theta_2) \ .
 \label{a12}
\eeq
 This boundary Yang-Baxter equation along with unitarity and crossing constraints can be used to find exact R-matrices \cite{Ghoshal:1994bc,Moriconi:2001xz}. As we see below, these solutions lie at distinguished points of the convex space of allowed R-matrices. 
 We emphasize that we do not assume integrability or particle number conservation in the numerical R-matrix bootstrap procedure we use later in the paper.  Let us now consider a few important integrable cases.
 
\subsubsection{NLSM, diagonal R-matrix}  
 In the case of the NLSM, there are various forms of the R-matrix that are known to be integrable \cite{Moriconi:2001xz}. 
 Here we consider only two. The first is a diagonal R-matrix of the form
\beq
 R =\mathrm{diag} \{\underbrace{R_1(\theta),\ldots R_1(\theta)}_{k}, \underbrace{R_2(\theta) \ldots R_2(\theta)}_{N-k}\} \ ,
 \label{a13}
\eeq 
breaking the symmetry from $O(N)$ to $O(k)\times O(N-k)$.
The result is \cite{Ghoshal:1994bc}
\beqa
R_1(\theta) &=& -R_0(\theta) F_{\frac{\lambda+\pi}{2}}(\theta) F_{\frac{\lambda(N-k-1)+\pi}{2}}(\theta)\ , \\
R_2(\theta) &=& \frac{\frac{\lambda}{4}(N-2k)+i\theta}{\frac{\lambda}{4}(N-2k)-i\theta} R_1(\theta)\ , \\
R_0(\theta) &=& \frac{\Gamma(\frac12+\frac{\lambda}{4\pi}-i\frac{\theta}{2\pi}) \Gamma(1+i\frac{\theta}{2\pi})\Gamma(\frac34+\frac{\lambda}{4\pi}+i\frac{\theta}{2\pi})\Gamma(\frac14-i\frac{\theta}{2\pi})}     {\Gamma(\frac12+\frac{\lambda}{4\pi}+i\frac{\theta}{2\pi})\Gamma(1-i\frac{\theta}{2\pi})\Gamma(\frac34+\frac{\lambda}{4\pi}-i\frac{\theta}{2\pi})\Gamma(\frac14+i\frac{\theta}{2\pi})}\ ,
 \label{a14}
\eeqa
where $R_0(\theta)$ is an auxiliary function, $F_a(\theta)$ was defined in eq.(\ref{a7}) and $\lambda$ in eq.(\ref{a6}). The cases $k=0$ and $k=N-1$ were studied in \cite{Ghoshal:1994bc} and correspond to Neumann and Dirichlet boundary conditions for all $O(N)$ indices. The intermediate values of $k$ were studied in \cite{Moriconi:2001xz} and correspond to mixed boundary conditions.
\subsubsection{NLSM, block diagonal R-matrix}
When $N$ is even, another possibility is an R-matrix of the form \cite{Moriconi:2001xz}
\beq
 R = \left(\begin{array}{ccccc}
 A(\theta) & iB(\theta) & 0 & 0 & \cdots \\
 -iB(\theta) & A(\theta) & 0 & 0 & \cdots \\
 0 & 0 & A(\theta) & iB(\theta) & \cdots \\
 0 & 0 & -iB(\theta) & A(\theta) & \cdots \\
 \vdots &  \vdots & \vdots & \vdots & \ddots 
 \end{array}\right)\ ,
 \label{a15}
\eeq
namely $N/2$ equal anti-symmetric blocks along the diagonal. The symmetry is $U(\frac{N}{2})$. The functions $A(\theta)$ and $B(\theta)$ are real analytic. It is not our purpose to consider all integrable cases, see \cite{Aniceto:2017jor,Moriconi:2001xz} for a nice summary, but this particular one has a free parameter $\alpha$ and leads to a case where the boundary of the space corresponds to integrable solutions. The functions $A(\theta)$, $B(\theta)$ are given by
\beqa
 B(\theta) &=& -i\alpha \theta A(\theta)\ , \\
 A(\theta) &=& \frac{1}{1-i\alpha\theta}\ F_{\pi+\frac{1}{\alpha}}(\theta)\, R_0(\theta)\ .
 \label{a16}
\eeqa
\subsubsection{pYB, diagonal R-matrix}
Now we can consider the pYB solution\footnote{As far as we know there is no physical model description for this theory. Perhaps this reflection matrix can help elucidate more its properties. It might be interesting to construct a model for it, an approach could be to try something similar to \cite{Bercini:2018ysh}.}. To our knowledge, the reflection matrix was not previously worked out in this case. Interestingly, we found it first numerically using the bootstrap and then we used the YB equation to get the analytic answer.  We only consider the case of an R-matrix of the form
\beq
 R =\mathrm{diag} \{\underbrace{R_1(\theta),\ldots R_1(\theta)}_{k}, \underbrace{R_2(\theta) \ldots R_2(\theta)}_{N-k}\}\ ,
 \label{a17}
\eeq 
with $1\le k\le \frac{N}{2}$. The boundary condition breaks the symmetry from $O(N)$ to $O(k)\times O(N-k)$. Using the Yang-Baxter equation and the explicit form of the S-matrix for the pYB solution leads to 
\beq
 \frac{R_1(\theta)}{R_2(\theta)} = \chi_{12}(\theta)= \frac{\sinh(\frac{i(\theta_0-\theta)\nu}{\pi})\cosh(\frac{i(\theta_0+\theta)\nu}{\pi}-\alpha)}{\sinh(\frac{i(\theta_0+\theta)\nu}{\pi})\cosh(\frac{i(\theta_0-\theta)\nu}{\pi}-\alpha)}\ ,
 \label{a18}
\eeq
where we defined $\alpha\in\mathbb{R}$ such that $k=\cosh\nu(1-\tanh\nu\tanh\alpha)$. The constant $\theta_0\in i \mathbb{R}$ defines a one-parameter family of integrable R-matrices. It is easy to see that, given a point $\theta_1$ on the imaginary axis of the physical region, the ratio $\frac{R_1(\theta_1)}{R_2(\theta_1)}$ takes every possible value as we change $\theta_0 \in i\mathbb{R}$. To obtain the $R_{1,2}(\theta)$ functions themselves we use the ratio to write the crossing identity in the form
\beq
 R_1(i\pi-\theta) = \frac{h(\theta)}{h(i\pi-\theta)} S_+^{pYB}(2\theta-i\pi)\, R_1(\theta)\ ,
 \label{a19}
\eeq
with
\beq
 h(\theta) = \frac{\sinh(\frac{2i\theta\nu}{\pi})}{\sinh(\frac{i\nu}{\pi}(\theta-\theta_0))\sinh(\frac{i\nu}{\pi}(\theta+\theta_0)+\frac{i\pi}{2}-\alpha)}\ .
 \label{a20}
\eeq
We have to solve equation (\ref{a19}) together with the condition that $R_1(\theta)$ is real analytic and saturates unitarity on the real axis $R_1(\theta) R_1(-\theta)=1$. It is useful to notice that if, for a moment, we ignore $S_+^{YB}(2\theta-i\pi)$ the general solution to eq.(\ref{a19}), up to CDD factors \cite{CDD}, is
\beq
 \hat{R}_1(\theta) = \prod_{j=0}^\infty \frac{h(-\theta-2\pi i j)h(-i\pi+\theta-2\pi i j)}{h(-\theta+i\pi-2\pi i j)h(-2 i\pi+\theta-2\pi i j)}\ .
 \label{a21}
\eeq 
In fact $S^{YB}_+(2\theta-i\pi)$ can be absorbed in $h(\theta)$ and then compute $R_1(\theta)$ with the result (up to the CDD factors described below)
\begin{align}
 R_1(\theta) &= \prod_{a=1}^{4}\, H_{\nu_a}\left(\frac{2i\theta\nu_a}{\pi}+\alpha_a,-\frac{2i\theta\nu_a}{\pi}+\beta_a;-\frac{2i\theta\nu_a}{\pi}+\alpha_a,\frac{2i\theta\nu_a}{\pi}+\beta_a\right) \ ,\label{a22}\\
 R_2(\theta) &= \frac{R_1(\theta)}{\chi_{12}(\theta)} \ , \label{a22b}
\end{align}
with $\chi_{12}(\theta)$ as in eq.(\ref{a18}), the function $H_\nu$ as defined in appendix A, and 
\begin{align}
\nu_1 &= \nu,\ &\alpha_1 &=3\nu ,\  &\beta_1 &=2\nu, \\
\nu_2 &= \nu,\ &\alpha_2 &=2\nu , &\beta_2 &=0, \\
\nu_3 &= \frac{\nu}{2},\ &\alpha_3 &=-\frac{i\theta_0\nu}{\pi}, &\beta_3 &=\nu-\frac{i\theta_0\nu}{\pi}, \\
\nu_4 &= \frac{\nu}{2},\ &\alpha_4 &=\frac{i\theta_0\nu}{\pi}+\frac{i\pi}{2}-\alpha,  &\beta_4 &=\nu+\frac{i\theta_0\nu}{\pi}+\frac{i\pi}{2}-\alpha.
 \label{a23}
\end{align}
It is also straightforward to use the identities in appendix \ref{H_nu} to check that this function solves eq.(\ref{a19}).  One can also check that the reflection matrix has the same periodicity as the S-matrix: $\theta\rightarrow\theta+\frac{2\pi^2}{\nu}$. However this is not necessarily a minimal solution. We consider a minimal solution one with no poles on the physical region $0\le\Im\theta\le \frac{\pi}{2}$ and also with no common zeros of $R_{1,2}$ in that region. Whether there are such zeros or poles depends on the value of the parameter $\theta_0$. With a bit of effort one can see that, if we define $\theta_0=2\pi i \xi_0$ and a periodic CDD factor\footnote{The period of the solution is $\frac{2\pi^2}{\nu}$ but the function changes sign when shifted by a half-period. Thus the periodicity of the zeros and poles is $\frac{\pi^2}{\nu}$.} as
\beq
 f_P(\alpha,\theta)= \prod_{n=-\infty}^{\infty} \frac{\sinh(\theta+\frac{n\pi^2}{\nu})-i\sin\alpha}{\sinh(\theta+\frac{n\pi^2}{\nu})+i\sin\alpha}\ ,
 \label{a24}
\eeq 
then we need to multiply $R_{1,2}(\theta)$ by one or both of the following CDD factors
\begin{itemize}
\item if $\xi_0<0$ then $\alpha=2\pi\left(\left\{\frac{1}{4}-\xi_0\right\}-\frac{1}{4}\right)$.
\item if $\xi_0>-\frac{\alpha}{2\nu}$ then  $\alpha=2\pi\left(\left\{\xi_0+\frac{\alpha}{2\nu}+\frac{1}{4}\right\}-\frac{1}{4}\right)$.
\end{itemize}
Here we used the notation $\{x\}=x-[x]$ where $[x]$ is the largest integer smaller or equal to $x$.
For a given value of $\theta_1$ we can plot the curve $(R_1(\theta_1,\theta_0=i\xi_0), R_2(\theta_1,\theta_0=i\xi_0))$ parameterized by $\xi_0\in\mathbb{R}$. That curve determines the boundary of the allowed values of $R_1(\theta_1)$, $R_2(\theta_1)$ and is depicted in fig.\ref{pYB_allowed_region} for $N=6$ and $k=1,2,3$. It has vertices where a new CDD factor comes in, namely for $\xi_0=0$ and $\xi_0=-\frac{\alpha}{2\nu}$.

\section{Numerics}

For the numerical bootstrap, we need to parameterize analytic functions on the strip $b_1 \le \Im\theta \le b_2$ for some real $b_{1,2}$. In principle, we only need the physical region corresponding to $b_1=0$, $b_2=\frac{\pi}{2}$ but it also turns out to be useful to consider other values of $b_{1,2}$. In addition, we impose an extra periodicity along the real axis $\theta\equiv\theta+2\omega$ for some $\omega$ that we choose. This periodicity is introduced to facilitate the numerics, however, some R-matrices (pYB) are actually periodic. Therefore, we are considering analytic functions on a cylinder that can be parameterized in terms of Fourier coefficients or in terms of the values of the real part at the boundary on a set of equally spaced points. We describe more details in appendix \ref{numerics_appendix}. The numerical procedure gives excellent results on the imaginary axis and, on the real axis on a region roughly $-\omega/2\lesssim \Re\theta \lesssim \omega/2$ since beyond that, boundary effects can be important. In the plots, we plot this central region where the solution is smooth. Although initially, we make all the plots for functions analytic in the physical region  $b_1=0$, $b_2=\frac{\pi}{2}$, we found that the allowed space of R-matrices drastically reduces if we extend the region of analyticity. Also, an integrable theory which was originally at a regular point of the boundary, now appears at a vertex of the reduced region. This also shows that all the excluded points corresponded to R-matrices with at least one pole in the extended region. This is an interesting way to determine where different R-matrices have poles.

\subsection{NLSM, diagonal R-matrix}
\begin{figure}
	\centering
    \includegraphics[width=.9\textwidth]{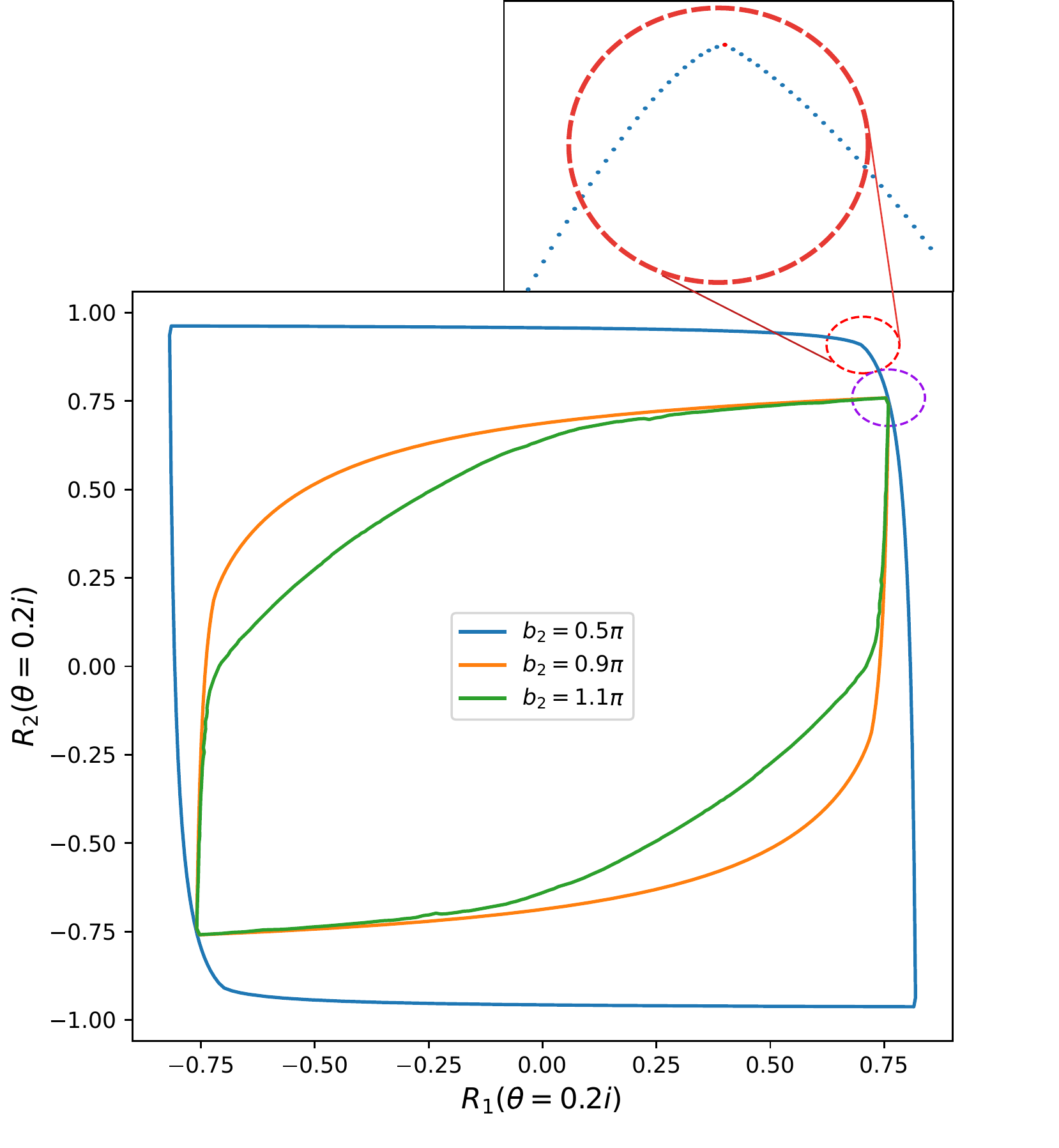}
	\caption{Plot of the $(R_1(\theta_1),R_2(\theta_1))$, $\theta_1=0.2i$, allowed region for the NLSM with $N=6$, $k=1$ and various analytic regions $b_1=0$, $b_2=\frac{\pi}{2}, 0.9\pi$, and $b_1=-0.1\pi$, $b_2=1.1\pi$. The vertex at the center of the red circle (see rotated and rescaled inset) is the R-matrix in eq.(\ref{a14}) with $N=6$, $k=1$. A new vertex appears in the smaller regions, namely the one at the center of the purple circle that corresponds to a solution with $R_1=R_2$, or $N=6$, $k=0$.}
	\label{NLSM_allowed_regions}
\end{figure}

\begin{figure}
	\centering
	\subfloat[$R_1(\theta)$, Dirichlet]{\includegraphics[width=.52\textwidth]{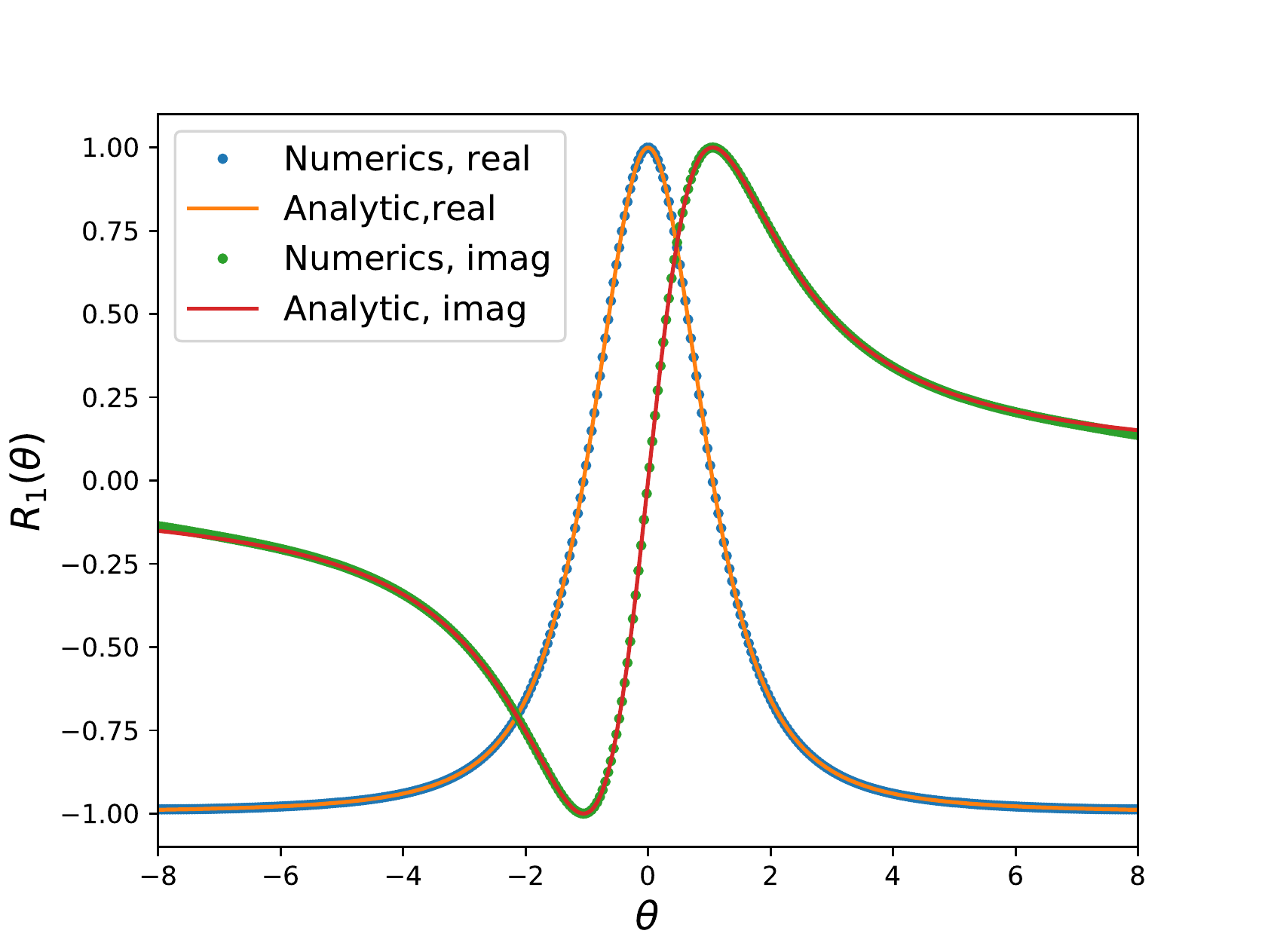}}
  \subfloat[$R_2(\theta)$, Dirichlet]{\includegraphics[width=.52\textwidth]{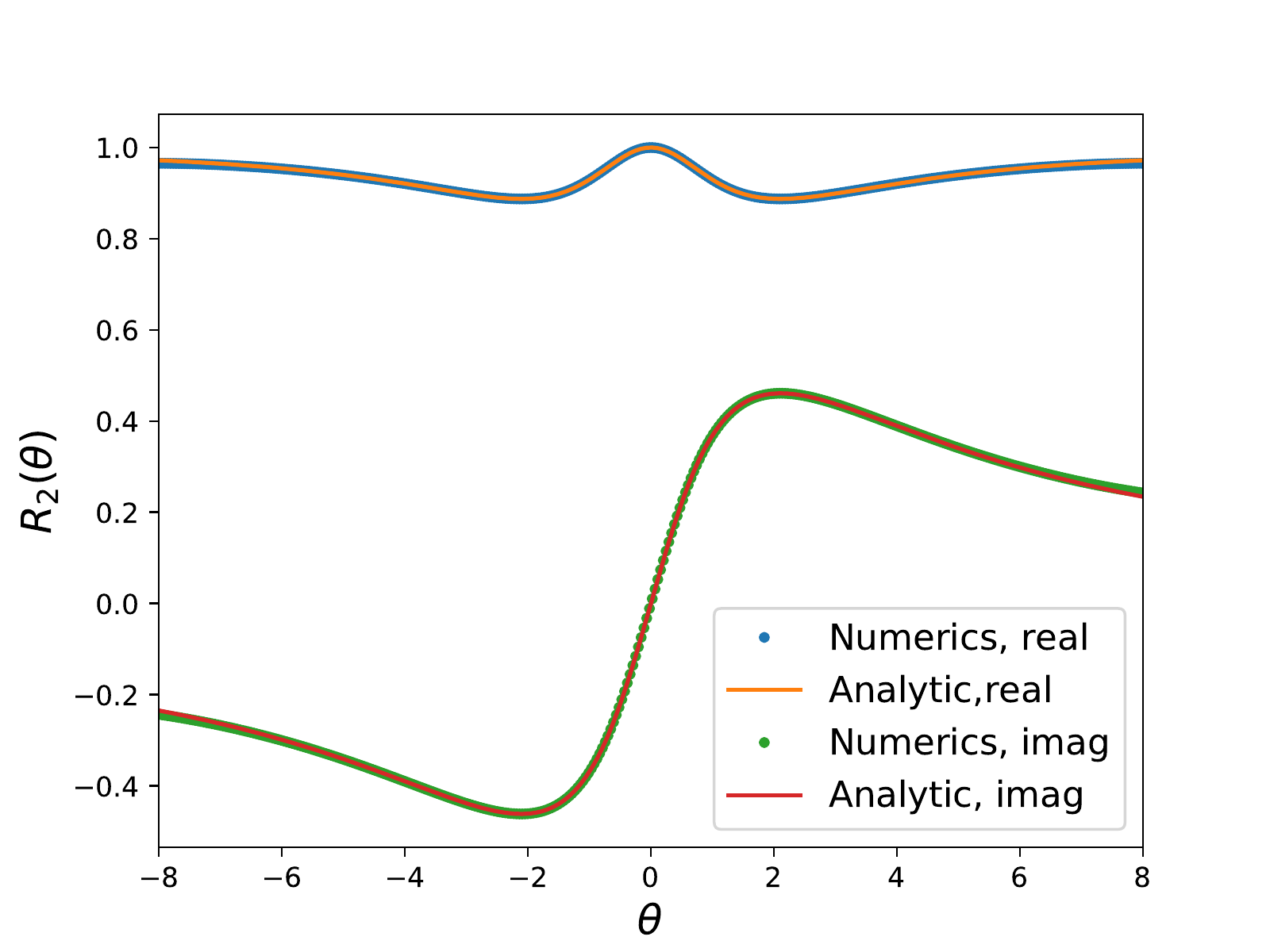}}\\
  \subfloat[$R_1(\theta)$, Neumann]{\includegraphics[width=.52\textwidth]{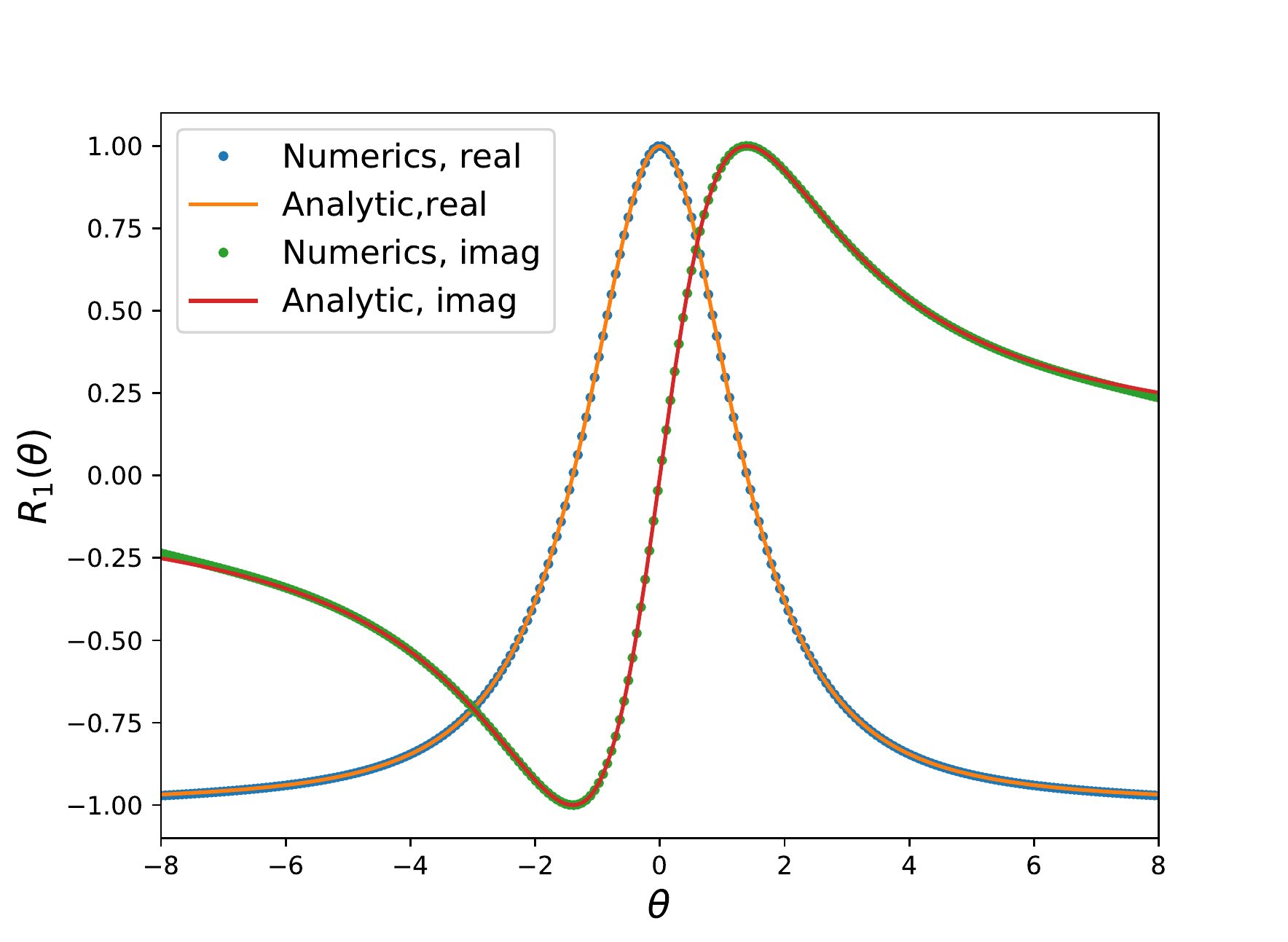}}
	\caption{Plot of $R_1(\theta)$ and $R_2(\theta)$ on the real axis. Figures (a) and (b) correspond to the red dashed vertex of fig.\ref{NLSM_allowed_regions}. It agrees precisely with Dirichlet R-matrix. Figure (c) is the purple dashed vertex of fig.\ref{NLSM_allowed_regions} and agrees with the Neumann R-matrix.}
	\label{NLSM_diag_Rmatrix}
\end{figure}

\begin{figure}
	\centering
	\subfloat[$k=2$]{\includegraphics[width=.5\textwidth]{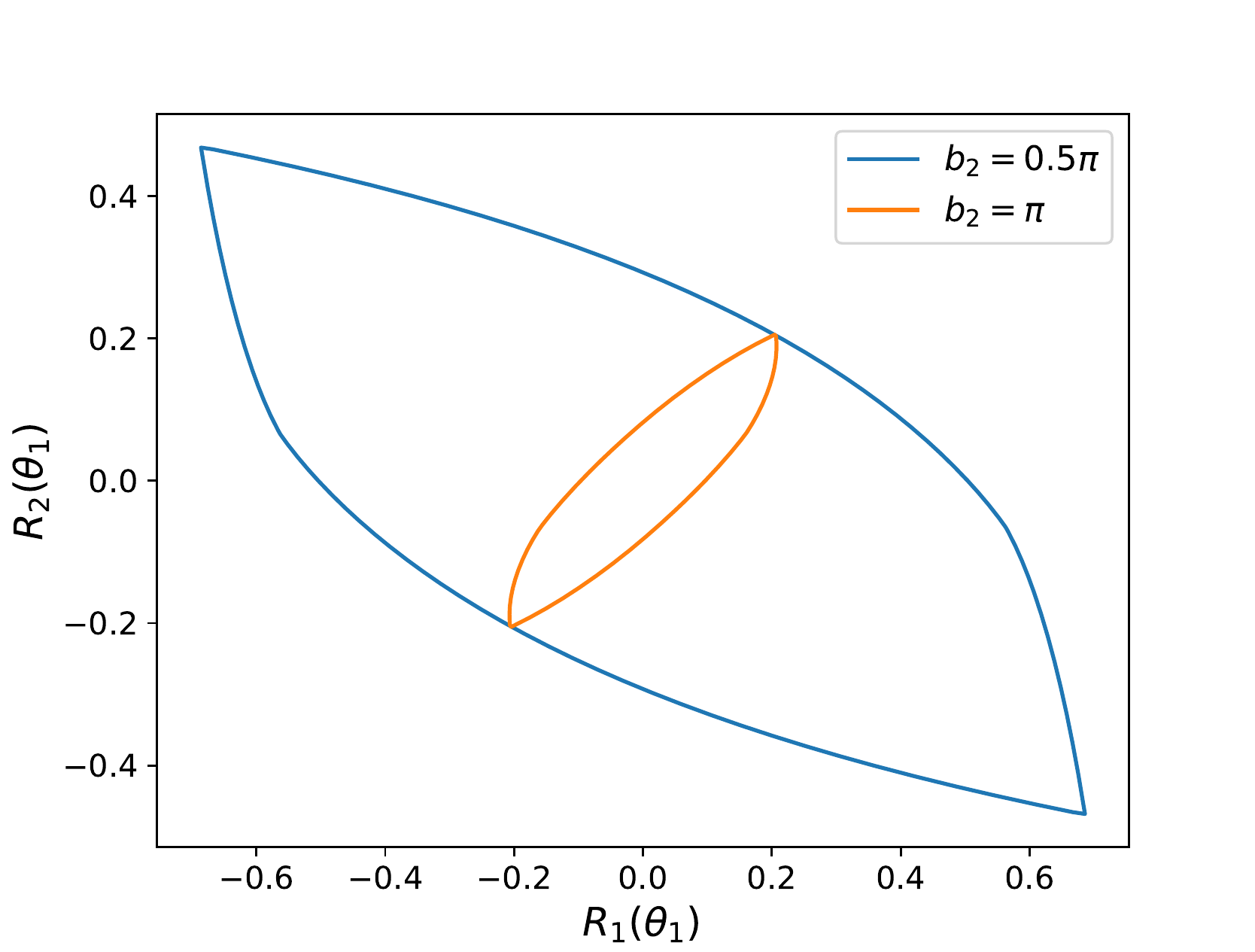}}
	\subfloat[$k=3$]{\includegraphics[width=.5\textwidth]{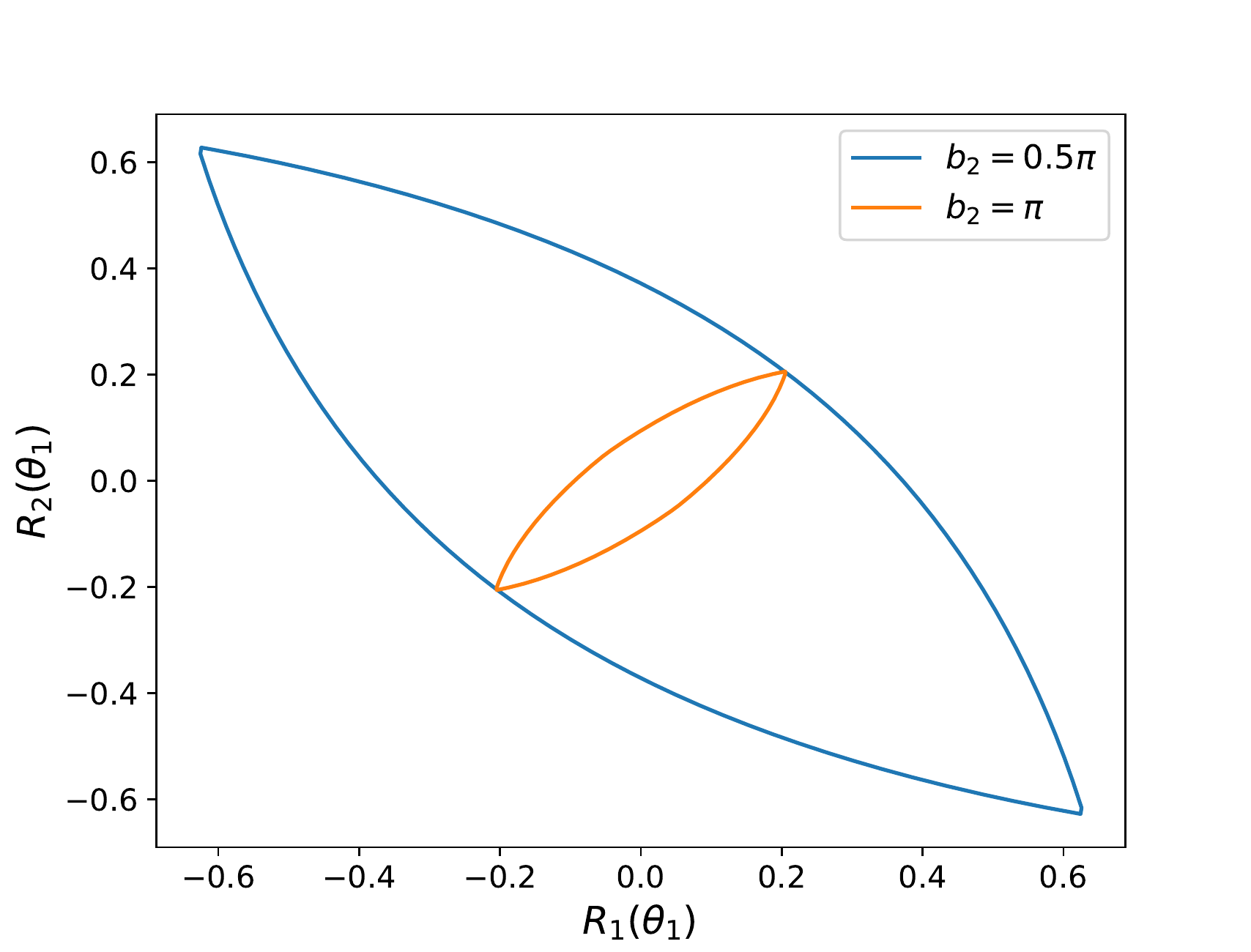}}
	\caption{Plot of the $(R_1(\theta_1),R_2(\theta_1))$ allowed region for the NLSM and $\theta_1=0.9879i$. We consider the case $N=6$, and $k=2,3$ and analyticity regions $b_1=0$, $b_2=\frac{\pi}{2},\pi$.}
	\label{NLSM_allowed_regions_k}
\end{figure}

Here we consider the bulk S-matrix to be given by the NLSM and assume the ansatz in eq.(\ref{a13}):
\beq
  R =\mathrm{diag} \{\underbrace{R_1(\theta),\ldots R_1(\theta)}_{k}, \underbrace{R_2(\theta) \ldots R_2(\theta)}_{N-k}\}\ .
 \label{a25}
\eeq
We then choose a point $\theta_1=i\xi_1$, $0<\xi_1<\frac{\pi}{2}$ and plot the boundary of the two dimensional region of allowed values of $(R_1(\theta_1), R_2(\theta_1))$. Notice that they are real by real analyticity. The results are plotted in fig.\ref{NLSM_allowed_regions} for the case $N=6$, $k=1$. The largest region corresponds to the allowed values of $(R_1(\theta_1), R_2(\theta_1))$ when imposing analyticity in the physical region. An integrable solution appears at a vertex indicated by the red circle corresponding to Dirichlet boundary conditions for one of the fields. The analytic form of the functions $R_{1,2}(\theta)$ is given in eq.(\ref{a14}) and is shown in fig.\ref{NLSM_diag_Rmatrix} to agree perfectly with the numerics. Another integrable solution exists for which $R_1(\theta)=R_2(\theta)$, the one corresponding to Neumann boundary conditions and given also by eq.(\ref{a14}) with $k=0$. It is at the center of the purple circle but it is not at a vertex of the curve. There is a clear vertex on the upper left part of the curve although we were not able to identify it with a known theory. To make the other integrable vertex manifest we  find now the allowed region for R-matrices analytic up to $0.9\pi$. This drastically reduces the allowed space to the region surrounded by the orange curve. The integrable solution now is clearly at a vertex!. Further increasing the region of analyticity to $1.1\pi$ gives the green curve that does not seem to display any new vertices. As can be seen in fig.\ref{NLSM_allowed_regions_k} the same phenomenon happens when taking $N=6$ and $k=2$ and $k=3$. 

\subsection{NLSM, block diagonal R-matrix}
\begin{figure}
	\centering
\includegraphics[width=.9\textwidth]{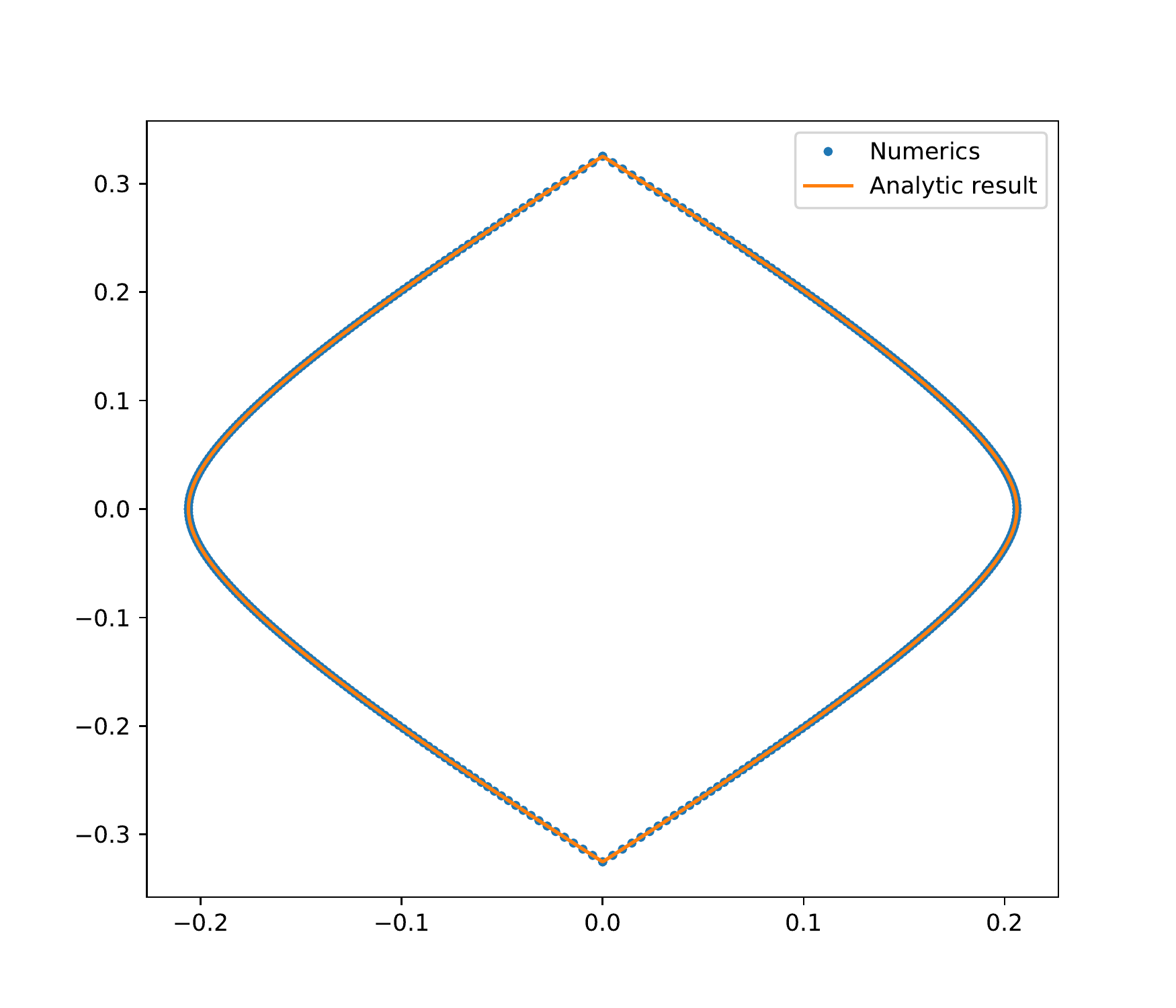}
	\caption{Allowed values of $A(\theta_1),B(\theta_1)$ for $\theta_1=0.9879\,i$. The solid curve is the analytic result and the points are the numerical data. All points of the curve correspond to the same integrable R-matrix that has a free parameter $\alpha$ that changes along the boundary, see eq.(\ref{a16}). The vertex corresponds to the limit $\alpha\rightarrow\infty$.}
\label{ABshape}
\end{figure}

\begin{figure}
	\centering
	\subfloat[$A(\theta)$ and $B(\theta)$ functions on the imaginary axis for $\alpha=0.5$]{\includegraphics[width=1\textwidth]{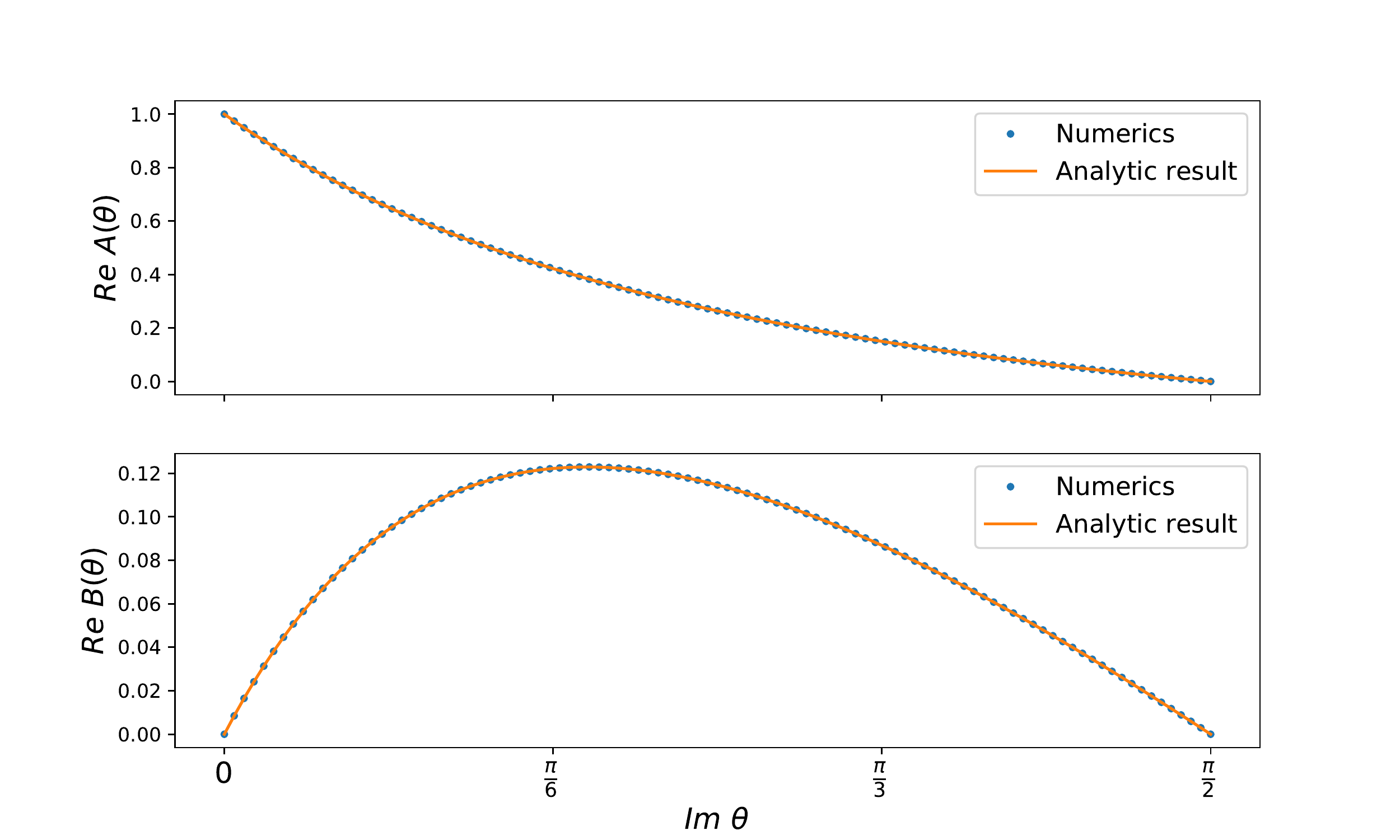}} \\
	\subfloat[$A(\theta)$ and $B(\theta)$ functions on the real axis for $\alpha=0.5$]{\includegraphics[width=1\textwidth]{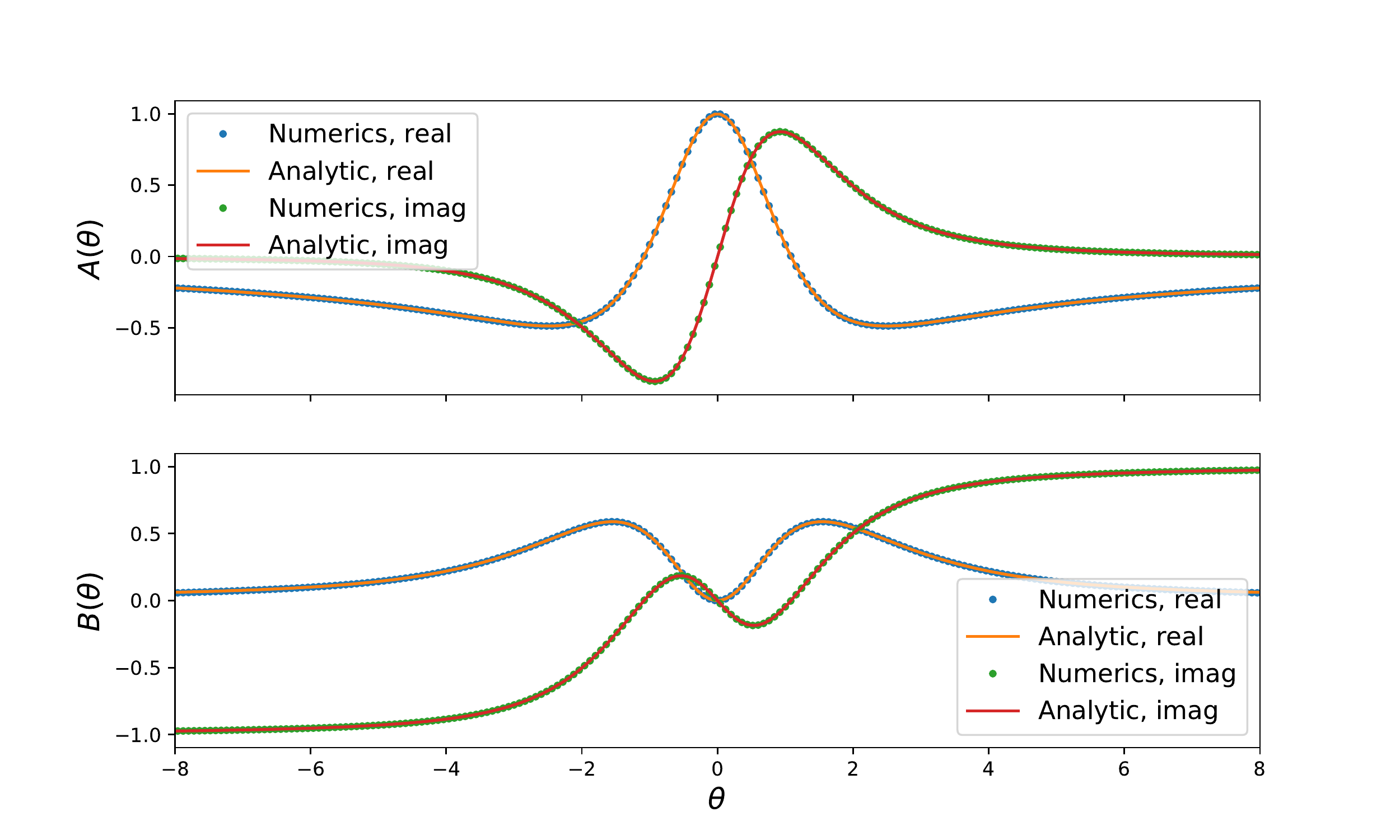}} 
	\caption{Plot of $A(\theta)$ and $B(\theta)$ along the real and imaginary axis of the physical region $0\le\Im \theta\le \frac{\pi}{2}$ for the off diagonal R-matrix ansatz. We find perfect agreement between the R-matrix bootstrap and the exact answer from integrability. }
	\label{AB_R-matrix}
\end{figure}

Here we consider the bulk S-matrix to be given by the NLSM and assume the ansatz in eq.(\ref{a15})
\beq
 R = \left(\begin{array}{ccccc}
 A(\theta) & iB(\theta) & 0 & 0 & \cdots \\
 -iB(\theta) & A(\theta) & 0 & 0 & \cdots \\
 0 & 0 & A(\theta) & iB(\theta) & \cdots \\
 0 & 0 & -iB(\theta) & A(\theta) & \cdots \\
 \vdots &  \vdots & \vdots & \vdots & \ddots 
 \end{array}\right)\ ,
 \label{a26}
\eeq
We then choose a point $\theta_1=i\xi_1$, $0<\xi_1<\frac{\pi}{2}$ and plot the boundary of the two dimensional region of allowed (real) values of $(A(\theta_1), B(\theta_1))$. Numerically we can verify that the relation $ B(\theta) =-i \alpha \theta A(\theta)$ is satisfied all around the boundary where the constant $\alpha$ depends on the boundary point. In fact we can plot the boundary curve using the analytic solution in perfect agreement with the numerical results, see fig.\ref{ABshape}. Moreover, we can check various points at the boundary and see that the R-matrix itself agrees as exemplified in fig.\ref{AB_R-matrix}.

\subsection{pYB, diagonal R-matrix}
\begin{figure}
	\centering
	\includegraphics[width=1.1\textwidth]{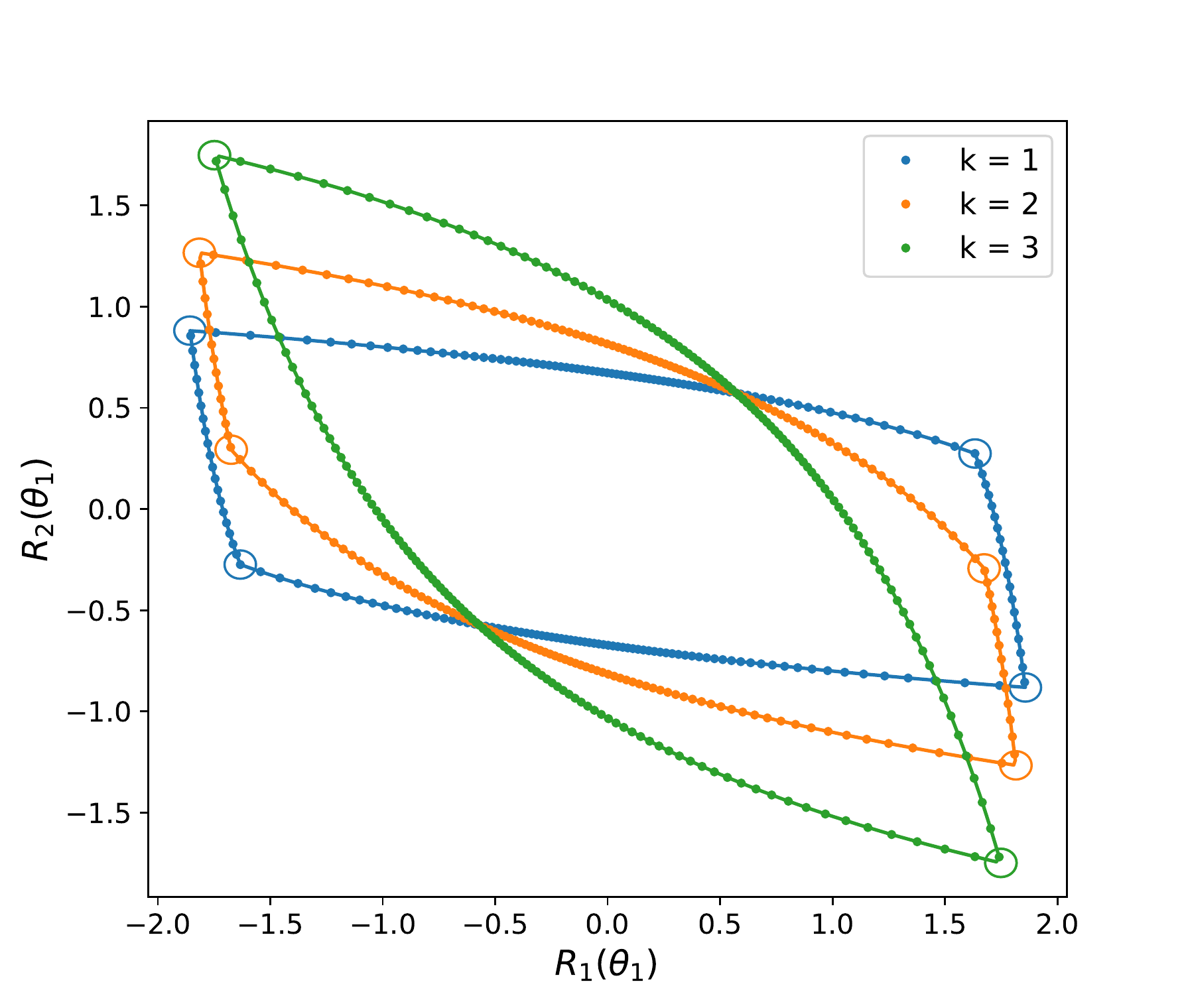}
	\caption{Plot of the $(R_1(\theta_1),R_2(\theta_1))$ allowed region for the pYB model for $\theta_1=0.9879i$. We consider the case $N=6$, and $k=1,2,3$. In this case, the whole boundary is integrable and we can compare the numerics (dots) with the analytical result (solid lines). The curves intersect at $R_1(\theta)=R_2(\theta)$ common to all $k$ in the ansatz (\ref{a27}). The vertices (encircled) appear when a pole or zero enters that physical region and has to be cancelled by changing the CDD factor.}
	\label{pYB_allowed_region}
\end{figure}

\begin{figure}
	\centering
	\subfloat[$N=6$, $k=1$]{\includegraphics[height=0.27\textheight]{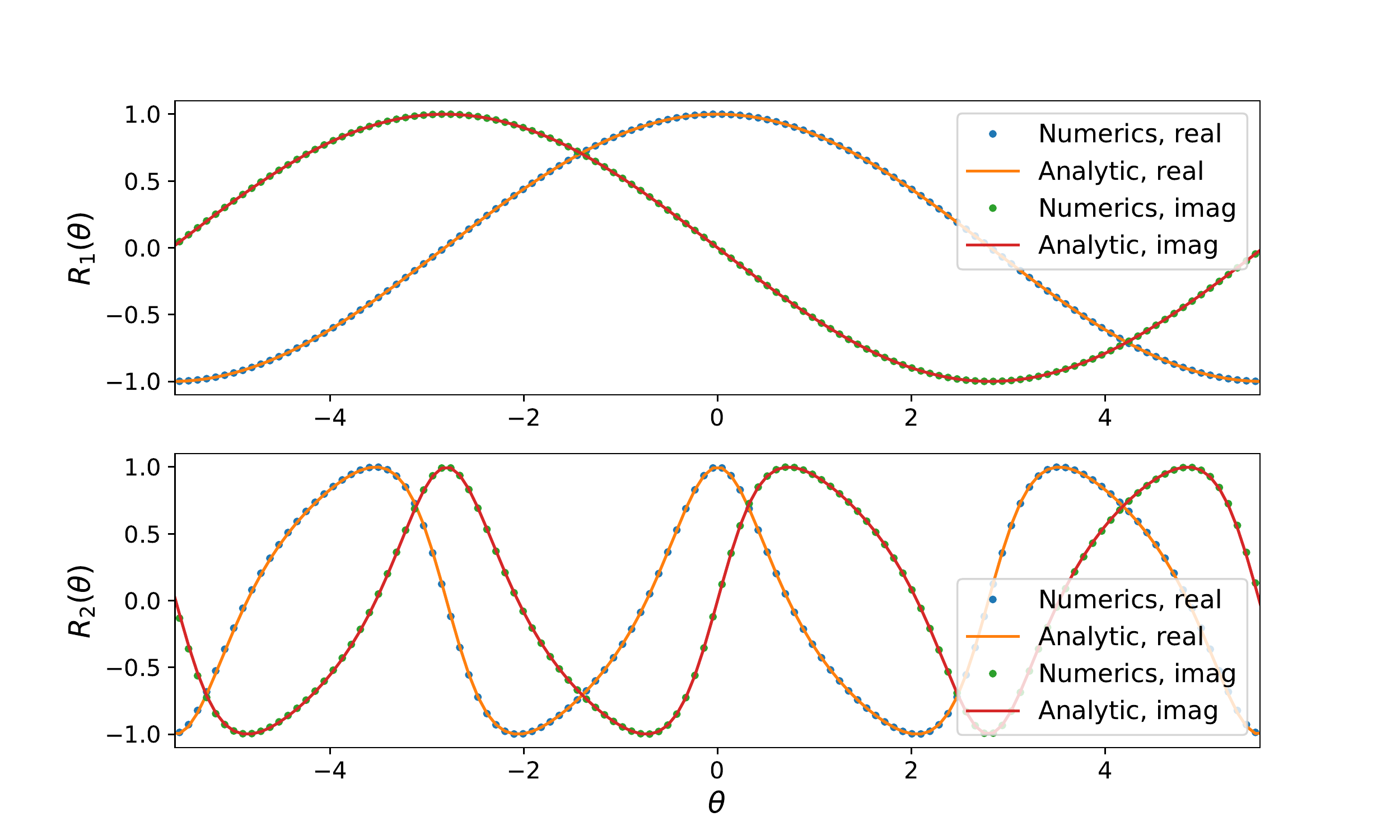}} \\
	\subfloat[$N=6$, $k=2$]{\includegraphics[height=0.27\textheight]{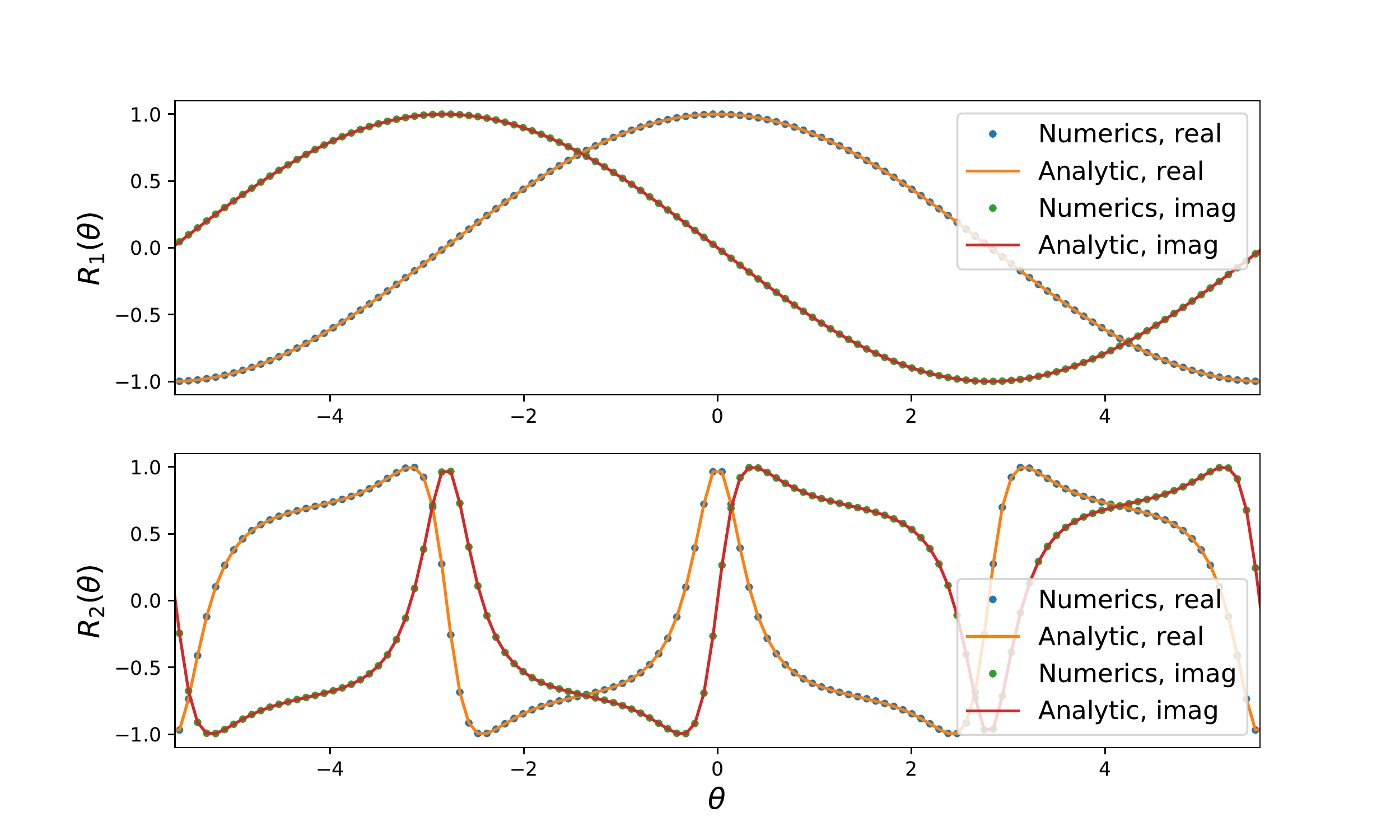}} \\
	\subfloat[$N=6$, $k=3$]{\includegraphics[height=0.27\textheight]{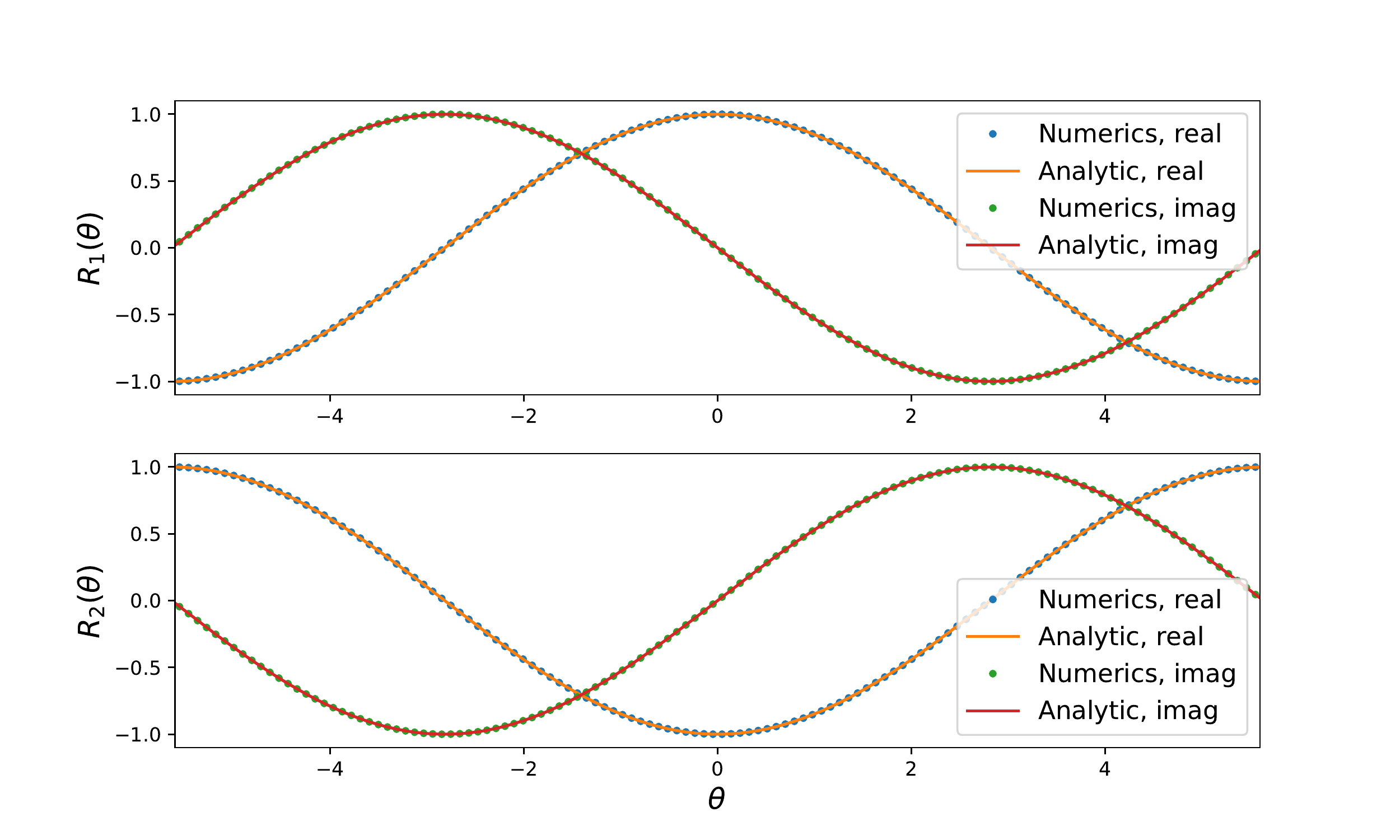}}
	\caption{Real and imaginary part of $R_1(\theta)$ and $R_2(\theta)$  on the real axis. There is perfect agreement between the R-matrix bootstrap and the analytic solutions. We take $N=6$, $k=1,2,3$, and $\theta_0=i\frac{\pi}{4}, -0.329i,0$.}
	\label{pYB_R-matrix}
\end{figure}

Here we consider the bulk S-matrix to be given by the pYB and assume the ansatz in eq.(\ref{a17})
\beq
  R =\mathrm{diag} \{\underbrace{R_1(\theta),\ldots R_1(\theta)}_{k}, \underbrace{R_2(\theta) \ldots R_2(\theta)}_{N-k}\}\ .
 \label{a27}
\eeq
We then choose a point $\theta_1=i\xi_1$, $0<\xi_1<\frac{\pi}{2}$ and plot the boundary of the two dimensional region of allowed (real) values of $(R_1(\theta_1), R_2(\theta_1))$. 
Again, there is a one-parameter family of analytic R-matrices and we can plot (see fig.\ref{pYB_allowed_region}) the allowed region analytically in perfect agreement with the numerical one. For various points at the boundary there is also good agreement between the numerical results and the exact ones eqs.(\ref{a22}), (\ref{a23}) as seen in fig.\ref{pYB_R-matrix}. To find agreement it is important to include the CDD factor of eq.(\ref{a24}). The required CDD factor depends of the point at the bounday. In this case, the vertices of the shape correspond to the points where the CDD factor changes, namely $\xi_0=0,-\frac{\alpha}{2\nu}$. In the case $N=6$, $k=3$ we have $\alpha=0$ and therefore there is only one vertex. That the CDD factor changes means that, at a vertex, a pole (or a zero) is moving into the physical region.

\section{The dual problem}
In this section, we discuss the dual problem. For the physical region, it works along the lines found for the S-matrix. When considering a region of extended analyticity some new results are found that are also applicable to the S-matrix. We do not provide numerical results since they all agree with the primal problem.

\subsection{Dual problem in the physical region}
The allowed space of R-matrices can be identified by maximizing various linear functionals while imposing crossing and unitarity constraints over linear combinations of a basis of functions. As we increase the basis, we cover a larger region and approach the boundary from the inside. We can also formulate a dual problem where we exclude points that do not satisfy the constraints by using a set of testing functions that partially impose the constraints. As we increase the set of functions, more constraints are imposed and we contract the region ending up with the same allowed region \cite{Cordova:2019lot}. Let the functional we wish to maximize in the primal problem be 
\beq
\mathcal{F}_P = \Re\left[n_a^b R_b^a(\theta_1)\right]\ ,
 \label{a28}
\eeq
for some constant real coefficients $n_a^b$ and a given point $\theta_1$ usually taken on the imaginary axis in which case taking real part is redundant due to real analyticity. Consider now a set of dual analytic functions $K_a^b(\theta)$ with a simple pole at $\theta=\theta_1$ and residue $n_a^b$ and perform the following integral along the real axis:
\beq
\mathcal F = \Re\left[\frac 1{2\pi i}\int_{-\infty}^{+\infty} d\theta K_a^b(\theta)R_b^a(\theta)\right]\ .
 \label{a29}
\eeq
 Now we move the contour of integration from the real axis to the line $\Im\theta=\frac{\pi}{2}$. Taking into account that we cross the pole we get
\beq
 \mathcal F = \Re\left[n_a^b R_b^a(\theta_1)\right]+ \Re\left[\frac 1{2\pi i}\int_{-\infty}^{+\infty} d\theta K_a^b(\frac{i\pi}{2}+\theta)R_b^a(\frac{i\pi}{2}+\theta)\right]\ .
 \label{a30}
\eeq
If we now impose anti-crossing condition for $K_i^j(\theta)$ 
\beq
K_a^b(i\frac\pi2-\theta) = -S_{ab}^{cd}(-2\theta)K_c^d(i\frac\pi 2+\theta)\ ,
\label{anti_cross_K}
\eeq
then we can change variable $\theta\rightarrow-\theta$ in the integral and use crossing (\ref{cross_unit}) to get
\beq
 \mathcal F = \Re\left[n_a^b R_b^a(\theta_1)\right] - \Re\left[\frac 1{2\pi i}\int_{-\infty}^{+\infty} d\theta K_a^b(\frac{i\pi}{2}+\theta)R_b^a(\frac{i\pi}{2}+\theta)\right]\ .
 \label{a31}
\eeq
Adding eqs.(\ref{a30}) and (\ref{a31}) we get $\cF=\cF_P$, namely,
\beq
 \mathcal{F}_P = \Re\left[n_a^b R_b^a(\theta_1)\right] =  \Re\left[\frac 1{2\pi i}\int_{-\infty}^{+\infty} d\theta K_a^b(\theta)R_b^a(\theta)\right]\ .
 \label{a32}
\eeq
Putting a bound on the right hand side of eq.(\ref{a32}) is easy if $R_a^b$ and then $K_a^b$ are diagonal as in eq.(\ref{a13}) since we just use that the diagonal elements of $R$ have modulus less than one. If we want to be more generic we can use that\footnote{This can be easily proven using the singular value decomposition $K=U\Sigma V^\dagger$ where $U$, $V$ are unitary and $\Sigma$ is diagonal with non-negative real entries}
\beq
  \left(\begin{array}{cc}\sqrt{K^\dagger K} & -K^\dagger \\ -K & \sqrt{K K^\dagger} \end{array}\right) \succeq 0\ ,
\eeq
and that the unitarity constraint can be written as
\beq
 \left(\begin{array}{cc}\mathbb{I} & R \\ R^\dagger & \mathbb{I} \end{array}\right) \succeq 0\ .
\eeq
Taking into account that, if $A$ and $B$ are positive semi-defininte then $\tr AB\ge0$ we get
\beq
 \half \tr \left[  \left(\begin{array}{cc}\sqrt{K^\dagger K} & -K^\dagger \\ -K & \sqrt{K K^\dagger} \end{array}\right)  \left(\begin{array}{cc}\mathbb{I} & R \\ R^\dagger & \mathbb{I} \end{array}\right) \right] = \tr \sqrt{K K^\dagger} +\half\tr(K^\dagger R^\dagger+KR) \ge 0\ ,
\eeq
to derive 
\beq
\mathcal{F}_P \le \frac{1}{2\pi} \int_{-\infty}^\infty d\theta\ \sum_{a=1}^N k_a(\theta) =\cF_D\ ,
 \label{a33}
\eeq
where $k_a(\theta)$, $a=1\ldots N$ are the positive square root of the eigenvalues of $\mathbb{K}_{ab}(\theta)=K_a^c(\theta) \bar{K}_b^c(\theta)$. This defines the dual functional $\cF_D$ and the dual problem is the minimization of $\cF_D$ subject to the anti-crossing constraint (\ref{anti_cross_K}). As mentioned, the simplest case is when $R_a^b(\theta)$ is diagonal in which case we take $K_a^b$ to also be diagonal and then $k_a$ is given by the diagonal elements as $k_a(\theta)=|K_a^a(\theta)|$ (no sum over $a$). If it is not diagonal then it is easier to introduce two hermitian matrices $\mathbb{Y}_{1,2}(\theta)$ and define the dual problem as
\beq
 \cF_D = \frac{1}{4\pi} \int_{-\infty}^\infty d\theta\ \tr(\mathbb{Y}_1(\theta)+\mathbb{Y}_2(\theta)) \ ,
 \label{a34}
\eeq
subject to the constraint
\beq
 \left(\begin{array}{cc}\mathbb{Y}_1(\theta) & -K^\dagger(\theta) \\ -K(\theta) & \mathbb{Y}_2(\theta) \end{array}\right) \succeq 0\ ,
 \label{a35}
\eeq
for all $\theta\in\mathbb{R}$. Here $K$ denotes the matrix $K_a^b$. When the duality gap closes we have (at each $\theta\in \mathbb{R}$)
\beq
 \left(\begin{array}{cc}\mathbb{Y}_1 & -K^\dagger \\ -K & \mathbb{Y}_2 \end{array}\right) \left(\begin{array}{cc}\mathbb{I} & R \\ R^\dagger & \mathbb{I} \end{array}\right) = 0\ .
 \label{a36}
\eeq
From here we get 
\beq
 \mathbb{Y}_1 = K^\dagger R^\dagger, \\ \mathbb{Y}_2=K R,\ \ \ K(RR^\dagger-1)=0\ .
 \label{a37}
\eeq
The last equality implies that unitarity is saturated $RR^\dagger=1$ or, otherwise, $K$ is not invertible, \ie\ $KK^\dagger$ has at least one zero eigenvalue. If we assume that $K$ is invertible then unitarity is saturated and we can compute (using that $\mathbb{Y}_1$ is hermitian)
\beq
 R = (K^\dagger K)^{-\half}\, K^\dagger \ .
 \label{a38}
\eeq
 The inverse square root is well defined since $K^\dagger K$ is a positive definite matrix under the assumption that it is invertible. 
We  verified the previously obtained numerical results using the dual formulation and the duality gap is indeed zero as expected.

\subsection{Dual to the extended analyticity problem}
The dual problem for extended analyticity is quite interesting, and it allows for unitarity non saturation. Extending the analyticity region for the primal problem leads to more freedom in the dual functions and thus a smaller allowed region.

\begin{figure}
	\centering
	\includegraphics[width=.75\textwidth]{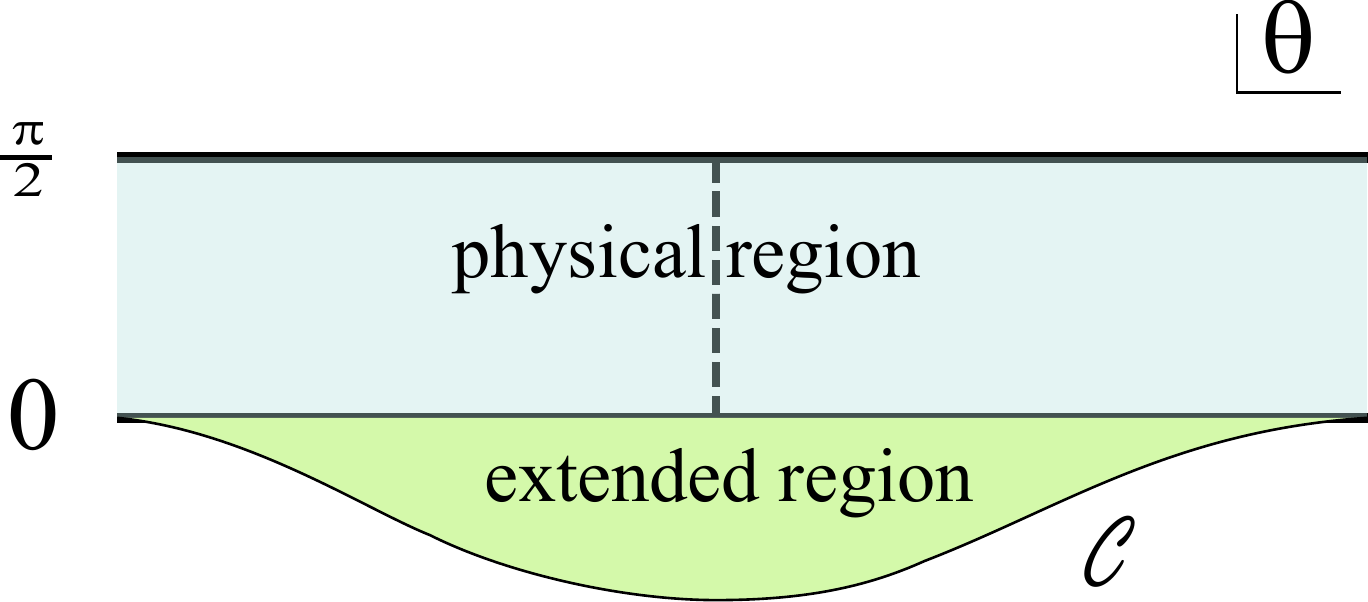}
	\caption{Extended region of analyticity. We generically extend the region below the real axis (across the physical cut) to a curve $\cC$. A well defined dual problem requires a regularization, namely a bound on the analytic function on $\cC$. An example of such curve is in fig.\ref{theta_extended_region} in which case the curve $\cC$ is $\Im \theta=b_1$. There we also extended the region to $\Im\,\theta = b_2$ which can be done in the same way.}
	\label{extended_analyticity}
\end{figure}

Once again, we rewrite the primal functional $\mathcal F_P=\Re\left[n_a^b R_a^b(\theta_1)\right]$ as an integral along the real axis imposing analyticity except for a pole at $\theta=\theta_1$ and also impose anti-crossing for $K_a^b(\theta)$, 
\beq
\mathcal F_P = \Re\left[\frac1{2\pi i} \int_{-\infty}^{+\infty} d\theta\ K_a^b(\theta)R_a^b(\theta)\right]\ .
 \label{a39}
\eeq
Now, we introduce a new function $\hat K_a^b(\theta)$ that is analytic in a region bounded by the real axis and a curve $\cC$ below it, see fig.\ref{extended_analyticity}. If we require the R-matrix $R_a^b(\theta)$ to also be analytic in that region we have the identity  
\beq
\frac1{2\pi i} \int_{\cC} \hat K_a^b(\theta)R_b^a(\theta) = \frac1{2\pi i}\int_{-\infty}^{+\infty} \hat K_a^b(\theta) R_b^a(\theta)\ ,
 \label{a40}
\eeq
allowing us to rewrite the primal functional as
\beqa
\mathcal F_P &=& \Re\left[\frac1{2\pi i} \int_{-\infty}^{+\infty} R_a^b(\theta)(K_b^a(\theta)-\hat K_b^a(\theta))\right] + \Re\left[\frac1{2\pi i} \int_{\mathcal C} \hat K_b^a(\theta)R_a^b(\theta) \right] \non\\
&\le& \frac1{2\pi} \int_{-\infty}^{+\infty} \sum_a \Delta k_a + \Re\left[\frac1{2\pi i} \int_{\mathcal C} \hat K_b^a(\theta)R_a^b(\theta) \right]\ ,
 \label{a41}
\eeqa
where, similar as before, $\Delta k_a$ are the positive square root of the eigenvalues of $\Delta K \Delta K^\dagger$ with $\Delta K_a^b=K_a^b-\hat{K}_a^b$. Unfortunately we do not have any bound on the last term in eq.(\ref{a41}). In fact, to have a well defined dual problem we need a regularization, namely we have to bound the values of $R_a^b(\theta)$ when $\theta$ is on the curve $\cC$. We can do that by using a bound similar to the unitarity bound
\beq
 RR^\dagger \preceq M \ ,
 \label{a42}
\eeq
where an identity matrix is assumed on the right hand side. With this in mind we write the dual functional as
\beq
 \cF_D = \frac1{2\pi} \int_{-\infty}^{\infty} \sum_a \Delta k_a + \frac{M}{2\pi} \int_{\mathcal C} \hat k_a(\theta) \left|\frac{d\theta(s)}{ds}\right| ds \ ,
  \label{a43}
\eeq
where $\theta(s)$ parameterizes the curve $\cC$ and $\hat{k}_a(\theta)$ are the positive square root of the eigenvalues of $\hat{K}(\theta(s)) \hat{K}^\dagger(\theta(s))$. Again $K(\theta(s))$ is the matrix $\hat{K}_a^b(\theta(s))$ on the curve $\cC$.    
 The dual problem can be conveniently described by taking $K$ and $\hat{K}$ as a single analytic function with a cut on the real axis. Then the first term in eq.(\ref{a43}) is an integral that measures the jump across the cut. When the duality gap closes we have the same analysis as before except that we should replace $K\rightarrow \Delta K$. If $\Delta K$ is invertible then unitarity is saturated. This is more easily expressed for a diagonal R-matrix and corresponding diagonal $\Delta K$. In that case, a diagonal element $R_a^a$ of the R-matrix does not have to saturate unitarity if the corresponding $\Delta K_a^a=0$. Then, in that region of the real axis, the functions $K$ and $\hat{K}$ are analytic continuations of each other. 

\section{Conclusions}
 
We presented an extension of the S-matrix bootstrap to the case of reflection matrices. We successfully reproduced known integrable R-matrices and found a new one which we then derived analytically. This shows that there is a new rich and interesting playground to test and develop S-matrix bootstrap ideas. In particular, a novel idea was the extended analyticity constraint. By requiring analyticity beyond the physical region, we contracted the allowed space of R-matrices, leading to a new vertex corresponding to an integrable model. This is a promising way to identify theories that can be readily applied to the S-matrix (work in progress). As regards the R-matrix bootstrap, we only explored a few possibilities and many other theories can be investigated including supersymmetric ones, theories with bound states, etc.

\section{Acknowledgments}

 We are very grateful to Luc\'\i a G\'omez C\'ordova, Yifei He, Nima Lashkari, and Pedro Vieira,  for comments and discussions.  We are also very grateful to the DOE that supported in part this work through grants DE-SC0007884, DE-SC0019202 and the QuantiSED Fermilab consortium, as well as to the Keck Foundation that also provided partial support for this work.

\appendix
\numberwithin{equation}{section}

\section{A useful function}
\label{H_nu}
 
Given four complex numbers $\alpha,\beta,\gamma,\delta$ with $\alpha+\beta=\gamma+\delta$ and a positive real number $\nu\in \mathbb{R}_{>0}$ we define\footnote{The equivalence of both definitions can be shown by checking that both have the same zeros and poles and their (quasi)-periodicity properties listed below, \eg\ behavior under $\alpha\rightarrow \alpha+4\nu$, etc.} 
\beqa
H_\nu(\alpha,\beta;\gamma,\delta) &=& \lim_{M\rightarrow\infty} \prod_{|n|\le M} \frac{\Gamma(\frac{\gamma+i\pi n}{4\nu})\Gamma(\frac{\delta+i\pi n}{4\nu})}{\Gamma(\frac{\alpha+i\pi n}{4\nu})\Gamma(\frac{\beta+i\pi n}{4\nu})}\\
&=& e^{\frac{\alpha\beta-\gamma\delta}{4\nu}} \lim_{M\rightarrow\infty}  \prod_{j=0}^{M}\frac{(e^{-2\alpha},e^{-8\nu})_\infty(e^{-2\beta},e^{-8\nu})_\infty}{(e^{-2\gamma},e^{-8\nu})_\infty(e^{-2\delta},e^{-8\nu})_\infty}\ ,
 \label{a44}
\eeqa
where $(a,q)_\infty=\prod_{j=0}^\infty (1-aq^j)$ is the q-Pochhammer symbol. The function $H_\nu$ can also be written as
\beqa
 H_\nu(\alpha,\beta;\gamma,\delta) &=& \prod_{n=-\infty}^{\infty} \frac{\Gamma(\frac{\gamma+i\pi n}{4\nu})\Gamma(\frac{\delta+i\pi n}{4\nu})}{\Gamma(\frac{\alpha+i\pi n}{4\nu})\Gamma(\frac{\beta+i\pi n}{4\nu})} \\
    &=& e^{\frac{\alpha\beta-\gamma\delta}{4\nu}} \prod_{j=0}^{\infty}\frac{\sinh(\alpha+4\nu j)\sinh(\beta+4\nu j)}{\sinh(\gamma+4\nu j)\sinh(\delta+4\nu j)}\ ,
 \label{a45}
\eeqa
with the understanding that the infinite products are computed as in the previous equation. This is an analytic function of $\alpha,\beta,\gamma,\delta$ on the whole complex plane except for poles whenever $\gamma=i n\pi -4\nu j$ or $\delta=in\pi-4\nu j$ for some $n\in \mathbb{Z}, j\in \mathbb{Z}_{\ge0}$. It also has zeros whenever $\alpha=in\pi -4\nu j$ or $\beta=in\pi-4\nu j$, $n\in \mathbb{Z}, j\in \mathbb{Z}_{\ge0}$. 
For values of $\alpha,\beta,\delta,\gamma$ other than the poles, convergence\footnote{This can be used to accelerate the convergence of the product of $\Gamma$ functions, namely expanding the $\log$ in inverse powers of $n$ and summing $\sum_{n=1}^\infty \frac{1}{n^\ell}=\zeta(\ell)$ for a few values of $\ell$.} is manifest if we take logs and expand the logs for large $n$ (in the case of $\Gamma$ using the Stirling approximation) while recalling that $\alpha+\beta=\gamma+\delta$.  To check crossing and unitarity of the solutions it is useful to notice the following symmetry, periodicity, quasi-periodicity and reality properties of this function:
\beqa
H_\nu(\alpha,\beta;\alpha,\beta)&=&1 \ ,\\
H_\nu(\alpha,\beta;\gamma,\delta)=H_\nu(\beta,\alpha;\gamma,\delta)&=&H_\nu(\alpha,\beta;\delta,\gamma)=\frac{1}{H_\nu(\gamma,\delta;\alpha,\beta)} \ ,\\
H_\nu(\alpha,\beta;\gamma,\delta)&=&H_\nu(\alpha,\beta;\sigma,\rho)H_\nu(\sigma,\rho;\gamma,\delta) \ ,\\
H_\nu(\alpha+i\pi,\beta+i\pi;\gamma+i\pi,\delta+i\pi)&=& H_\nu(\alpha,\beta;\gamma,\delta) \ ,\\
H_\nu(\alpha+i\pi,\beta-i\pi;\gamma,\delta)&=& e^{\frac{i\pi(\beta-\alpha)}{4\nu}+\frac{\pi^2}{4\nu}} H_\nu(\alpha,\beta;\gamma,\delta) \ ,\\
H_\nu(\alpha+i\pi,\beta;\gamma+i\pi,\delta)&=& e^{\frac{i\pi(\beta-\delta)}{4\nu}} H_\nu(\alpha,\beta;\gamma,\delta) \ ,\\
H_\nu(\alpha+4\nu,\beta;\gamma+4\nu,\delta)&=&  \frac{\sinh \gamma}{\sinh \alpha} H_\nu(\alpha,\beta;\gamma,\delta) \ ,\\
H_\nu(\alpha+4\nu,\beta-4\nu;\gamma,\delta)&=&  \frac{\sinh(\beta-4\nu)}{\sinh \alpha} H_\nu(\alpha,\beta;\gamma,\delta) \ ,\\
H_\nu(\alpha+4\nu,\beta+4\nu;\gamma+4\nu,\delta+4\nu)&=& \frac{\sinh \gamma \sinh \delta}{\sinh \alpha\sinh \beta} H_\nu(\alpha,\beta;\gamma,\delta) \ ,\\
(H_\nu(\alpha,\beta;\gamma,\delta))^* &=& H_\nu(\alpha^*,\beta^*;\gamma^*,\delta^*)\ ,
 \label{a46}
\eeqa  
and also the doubling identity
\beq
H_\nu(\alpha,\beta;\gamma,\delta) = H_{2\nu}(\alpha,\beta;\gamma,\delta)H_{2\nu}(\alpha+4\nu,\beta+4\nu;\gamma+4\nu,\delta+4\nu)\ ,
 \label{a47}
\eeq
that follows from $\Gamma(2x)=\frac{2^{2x-1}}{\sqrt{\pi}} \Gamma(x)\Gamma(x+\half)$. Using these identities we can define a function
\beq
 f_{a,b;\nu}(\theta)=H_\nu(a+\frac{2i\theta\nu}{\pi},b-\frac{2i\theta\nu}{\pi};a-\frac{2i\theta\nu}{\pi},b+\frac{2i\theta\nu}{\pi})\ ,
 \label{a48}
\eeq
such that
\beq
 f_{a,b;\nu}(i\pi-\theta)=H_{\nu/2}(a-\frac{2i\theta\nu}{\pi}-2\nu,b+\frac{2i\theta\nu}{\pi};a+\frac{2i\theta\nu}{\pi},b-\frac{2i\theta\nu}{\pi}-2\nu) f_{a,b;\nu}(\theta)\ .
 \label{a49}
\eeq
In particular
\beq
 f_{a,a+2\nu;\nu}(i\pi-\theta)=\frac{\sinh(a-\frac{2i\theta\nu}{\pi}-2\nu)}{\sinh(a+\frac{2i\theta\nu}{\pi})} f_{a,a+2\nu;\nu}(\theta)\ ,        
 \label{a50}
\eeq
can be used to solve the crossing equation (\ref{a19}) directly. 

\section{Discussion of numerical approach}
\label{numerics_appendix}
Consider functions $\Phi_a(\theta)$ analytic in the strip\footnote{In the main text we consider a strip $b_1\le\Im\theta\le b_2$ that can be obtained from here by a simple translation along the imaginary axis.} $0\le\Im \theta\le b$ and periodic along the real axis $\Phi_a(\theta+2\omega)=\Phi_a(\theta)$. In this case, they are the non-zero components of the R-matrix. Generically, the R-matrix is not periodic,  the periodicity is imposed to facilitate the numerics by cutting off the energy range. For numerical purposes, we parameterized (a subset) of such functions in two different, equivalent ways - (i) by the real parts of the boundary values of the function, and (ii) by the Fourier coefficients. We now discuss them both in brief.
\subsection*{Boundary values}

Let $v^a_j$, $\tilde{v}^a_j$ be the real part of the function $\Phi_a(\theta)$ at the points $\theta_j$ and $ib+\theta_j$ where
\beq
 \theta_j = -\omega+\frac{\omega}{M} (j-\half), \ \ \ j=1\ldots 2M \ ,
 \label{a51}
\eeq
We also define the functions 
\beqa
 h(\theta) &=& -\frac{1}{2M} +\frac{i}{2M} \frac{1-e^{\frac{i\pi M\theta}{\omega}}}{\tan\frac{\pi\theta}{2\omega}}+\frac{i}{M}\sum_{n=1}^{M}{}' \sin\frac{\pi n \theta}{\omega}\frac{e^{-\frac{\pi bn}{\omega}}}{\sinh\frac{\pi bn}{\omega}} \ ,\\
 \tilde{h}(\theta) &=& \frac{1}{2M} -\frac{i}{M} \sum_{n=1}^{M}{}' \frac{\sin\frac{\pi n \theta}{\omega}}{\sinh\frac{\pi bn}{\omega}} \ ,
 \label{a52}
\eeqa
where we used the notation $\sum'$ defined as
\beq
\sum_{n=1}^M{}'a_n =\sum_{n=1}^{M-1}a_n+\half a_M\ .
 \label{a53}
\eeq
Now the functions $\Phi_a(\theta)$ are taken to be
\beq
 \Phi_a(\theta) = \sum_{j=1}^{2M} v^a_j\, h(\theta-\theta_j) + \sum_{j=1}^{2M} \tilde{v}^a_j\,\tilde{h}(\theta-\theta_j)\ .
 \label{a54}
\eeq
It is easy to check that $v^a_j=\Re\Phi_a(\theta_j)$ and $\tilde{v}^a_j=\Re\Phi_a(ib+\theta_j)$. Notice that, for consistency, we need to impose
\beq
  \sum_{j=1}^{2M} v^a_j =  \sum_{j=1}^{2M} \tilde{v}^a_j
\eeq
which is equivalent to stating that the contour integral of the function around the domain is zero since there are no poles.

\subsection*{Fourier coefficients}
We can expand the function $\Phi_a(\theta)$ as
\beq
\Phi_a(\theta) = \sum_{n=-M}^{M} \a_n^a e^{\frac{in\pi\theta}{\omega}}\ ,
 \label{a55}
\eeq
where $M\in\mathbb{N}$ is a high frequency cut-off. Since we are working with real analytic functions, the coefficients $\a_n^a$ are real. Now, consider
\beq
\Phi_a(i b+\theta) = \a_0^a + \sum_{n=1}^{M} \a_n^a e^{-\frac{n\pi b}{\omega}}e^{\frac{in\pi\theta}{\omega}} + \a_{-n}^a e^{\frac{n\pi b}{\omega}}e^{-\frac{in\pi\theta}{\omega}}\ .
 \label{a56}
\eeq
For large $n$, the second term gets exponentially large if the coefficients $\a_{-n}^a$ are $\mathcal{O}(1)$. To avoid numerical issues while working with large numbers, we rescale the coefficients as
\beq
\tilde \a_{n}^a = \a_{-n}^a e^{-\frac{n\pi b}{\omega}}\ .
 \label{a57}
\eeq
With these new rescaled parameters, the function $\Phi_a(\theta)$ is
\beq
\Phi_a(\theta) = \a_0^a + \sum_{n=1}^{M} \a_n^a e^{\frac{in\pi\theta}{\omega}} + \tilde \a_{n}^a e^{-\frac{in\pi\theta}{\omega}}\ .
 \label{a58}
\eeq
In both cases, $M$ is typically taken from $40$ to $400$ depending on the accuracy needed. The period $\omega$ is taken large enough to determine the salient features of the functions taking into account that beyond $\Re\theta \sim\frac{\omega}{2}$ the boundary effects due to the imposed periodicity are notable. On the imaginary axis, the functions are determined very accurately.

In terms of either set of variables ($v^a_j, \tilde{v}^a_j$) or ($\a_0^a,\a_n^a,\text{and }\tilde\a_n^a$), the crossing constraints are linear (they can be imposed on the line $\Im\theta=\frac{\pi}{2}$ or between the upper line $\Im\,\theta=b$ and the line $\Im\,\theta=\pi-b$). The unitarity constraints are imposed at the points $\theta_j$ and are quadratic in the variables. The functional to maximize is taken as a linear function of the variables. Therefore the maximization problem becomes a standard conic convex optimization problem that can be solved by standard methods \cite{cvx}. For example we can take a point $\theta_1$ on the imaginary axis and 
impose
\beq
 R_1(\theta_1)=t\cos\xi,\ \ R_2(\theta_1) = t\sin\xi\ ,
 \label{a59}
\eeq 
and maximize $t$. By sweeping the values of $\xi\in[0,2\pi]$ and plotting the resulting $R_1(\theta_1), R_2(\theta_1)$ we find the boundary of the allowed region as depicted in figs.\ref{NLSM_allowed_regions}, \ref{NLSM_allowed_regions_k}, \ref{ABshape} and \ref{pYB_allowed_region}. Moreover, for any point at the boundary of the region, the functions $R_{1,2}(\theta)$ are determined with good accuracy except for boundary effects due to the imposed periodicity.

\section{Free bulk theory}
\label{free_theory_appendix}
 When the theory in the bulk is free, \ie\ $S_{ab}^{cd}=\delta_a^c\delta_b^d$, the boundary Yang--Baxter eq.(\ref{a12}) is trivially satisfied. In this appendix, we analytically obtain the allowed space of R-matrices with a free bulk and compare with numerics. The crossing equation for this case reads
 \beq
 R_a^b(i\pi-\theta) = R_b^a(\theta)\ .
 \eeq
 Once again, we consider diagonal (\ref{a13}) and block diagonal (\ref{a15}) reflections.
 
 For the diagonal case, we have two self-crossing functions that are bounded in the physical strip by modulus less or equal to one (by unitarity and crossing). So the allowed region for ($R_1(\theta_1)$, $R_2(\theta_1)$) should be contained in a square with vertices ($\pm 1, \pm 1$). These vertices satisfy all the constraints and so, the allowed region should contain them and by convexity also their convex hull, namely the said square. Therefore, the allowed region is the square with vertices at ($\pm 1, \pm 1$) corresponding to the usual Dirichlet/Neumann boundary conditions in different directions.

\begin{figure}
	 \centering
	\subfloat[Diagonal ansatz]{\includegraphics[height=0.26\textheight]{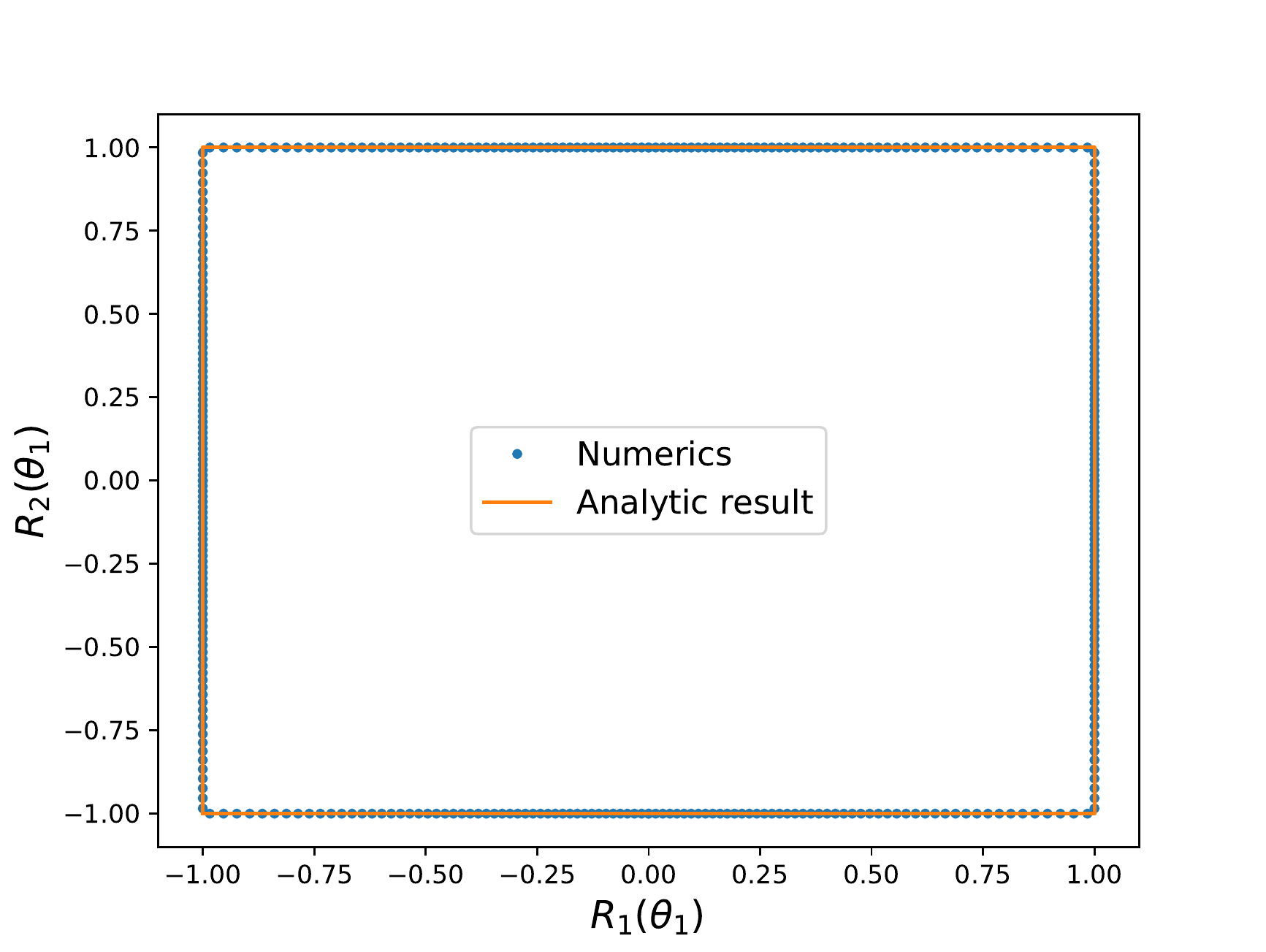}}
	\subfloat[Block diagonal ansatz]{\includegraphics[height=0.27\textheight]{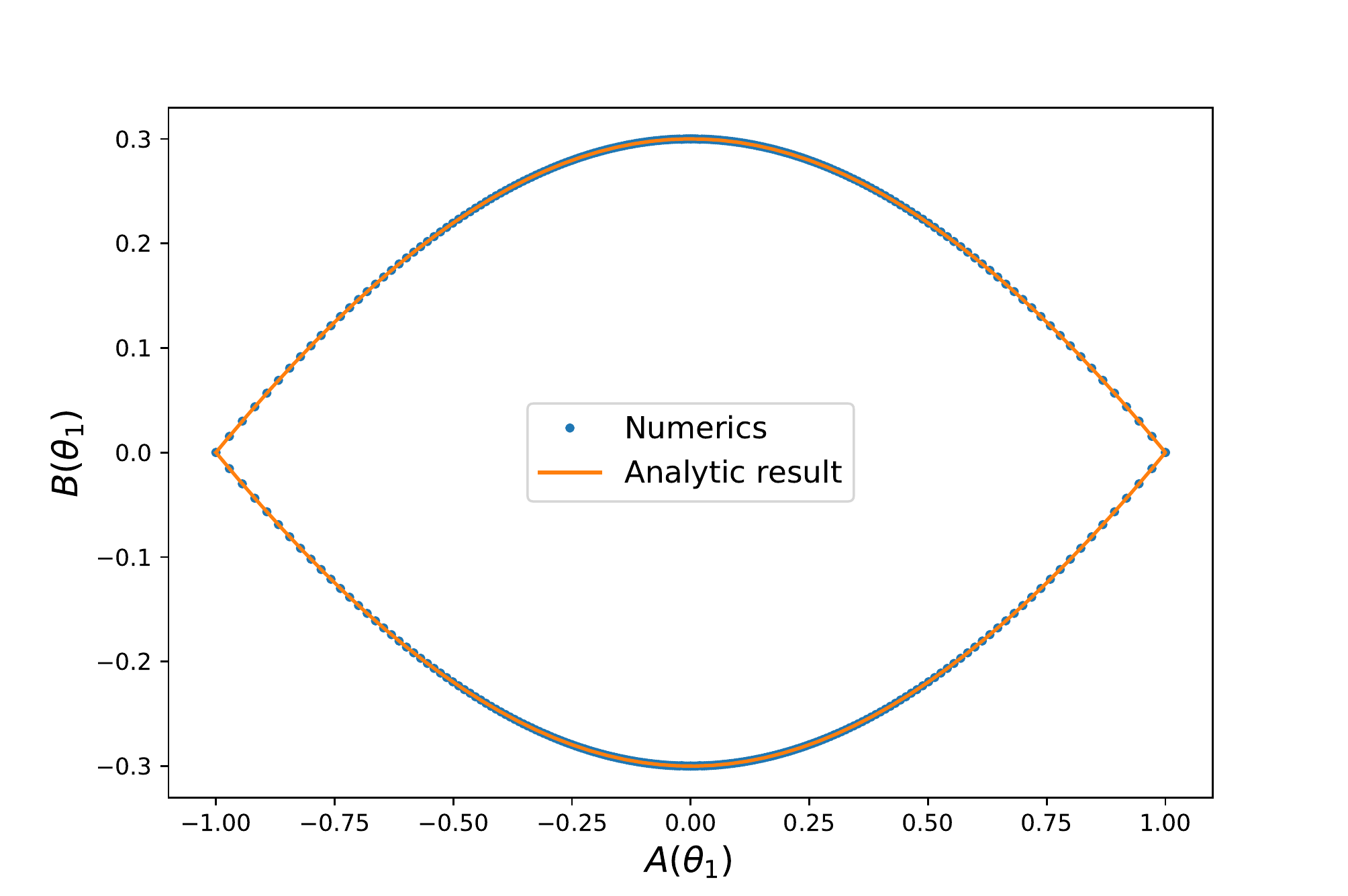}}
 	\caption{Allowed regions for a free bulk at $\theta_1 = 0.9879\,i$ in agreement with analytic results.}
 	\label{freeTheory}
 \end{figure}

 For the block diagonal case (\ref{a15}), the problem can be rewritten in terms of two analytic functions $f(\theta) = A(\theta) + B(\theta)$ and $g(\theta) = A(\theta) - B(\theta)$ as
 \beq
 \begin{aligned}
 \max_{f, g} \quad & \cos(\xi) f(\theta_1) + \sin(\xi) g(\theta_1) \ ,\\
 \textrm{s.t.} \quad & |f(\theta)| \le 1,\ \ |g(\theta)| \le 1\ \  \text{for}\ \  \theta \in \mathbb{R} \ ,\\
  					& f(i\pi-\theta) = g(\theta)   \ . \\
 \end{aligned}
 \eeq
 This problem can be solved analytically by mapping the strip $0\le\Im\ \theta\le\pi$ to a unit disk ($z = \frac{i-e^\theta}{i+e^\theta}$) and using the results from \cite{duren1970}. In the language of chapter 8 of \cite{duren1970}, the problem corresponds to a rational kernel with two poles ($n=2$) located at $z_1$ and $-z_1$ where $z_1=\frac{i-e^{\theta_1}}{i+e^{\theta_1}}$ and we are maximizing within the space $H_{p=\infty}$ of bounded analytic functions. Using the result below eq.(12), page 138 of \cite{duren1970} the extremal function takes the form
 \beq
 f(z) = \frac{iz + a}{1+iaz} \ ,
 \eeq
where we used that $n=2$ and $a$ is real by real analyticity. Here $-1\le a\le 1$ can be used as an arbitrary parameter that parameterizes the boundary of the allowed region which is then given by
 \beqa
 A(\theta) &=& \frac{2a \sinh(\theta)}{(a^2+1)\sinh(\theta)-i(a^2-1)}\ ,\\
 B(\theta) &=& \frac{i(a^2-1)\cosh(\theta)}{(a^2+1)\sinh(\theta)-i(a^2-1)}\ .
 \eeqa
 As a check, one can match these analytic results with the ones from the numerical bootstrap (see fig. \ref{freeTheory}). Although very simple, we are not aware of work where this off-diagonal free theory reflection matrix was further studied. 

\end{document}